\documentclass[preprint,12pt]{elsarticle}

\usepackage[english]{babel}

\usepackage{amsmath,amssymb,amsfonts}
\usepackage{graphicx}
\usepackage{dcolumn}
\usepackage{bm}
\usepackage[colorlinks=true, allcolors=blue]{hyperref}
\usepackage{enumitem}
\usepackage{indentfirst}

\usepackage{hyperref}
\hypersetup{colorlinks,allcolors=black}

\usepackage[export]{adjustbox}
\usepackage{comment}

\usepackage{lineno}

\newcommand{\ruru} {Ru+Ru}
\newcommand{\zrzr} {Zr+Zr}

\newcommand{\pt}   {\ensuremath{p_\mathrm{T}}}
\newcommand{\GeV}  {Ge\kern-.1emV}
\newcommand{\gevc} {GeV/$\textsl{c}$}
\newcommand{\snn}  {\ensuremath{\sqrt{s_{\mathrm{NN}}}}}

\begin{document}


\title{Study of event and particle selection effects on elliptic flow background at the isobar experiments based on AMPT model}

\author[1]{Yu Wang}
\author[1]{Hua Pei}
\affiliation[1]{
    organization={Key Laboratory of Quark and Lepton Physics (MOE) and Institute of Particle Physics},
    addressline={Central China Normal University}, 
    city={Wuhan},
    postcode={430079},
    country={China},
}

\cortext[cor1]{Corresponding author, \url{huapei@ccnu.edu.cn}}

\begin{abstract}

Measurement of the Chiral Magnetic Effect (CME) has been a popular topic of high-energy nuclear physics in the last decade.
The flow correlation $\gamma$ between charged hadron pairs of the same and opposite charges and their difference $\Delta \gamma$ were measured to separate the CME-driven signal from the collective flow background especially second-order elliptic $v_{2}$.
The STAR experiment have stepped further to the isobar experiment to compare $\gamma$ and $\Delta \gamma$ between \ruru\ and \zrzr~\cite{PhysRevC.105.014901}, which were theoretically expected to produce the same elliptic flow background but different CME signals.
However, the measured flow backgrounds also differ between \ruru\ and \zrzr\ , indicating more fine-tuning of RP and centrality definition necessary.

This analysis applied the AMPT model~\cite{PhysRevC.72.064901} to simulate the same collision system and energy as the STAR isobar experiment.
Since the AMPT model does not include magnetic field effects, we expect comparing its output between \ruru\ and \zrzr\ collision systems can provide an insight of the possible bias of flow background definition,
and help improve the measurement of CME signal in real experiments.
Multiple combinations of centrality and flow definition were chosen to study how the $v_2$ and their difference would be affected,
especially by varying the particles selection of charge versus neutral properties and broadening (pseudo-)rapidity regions, while STAR CME work relied on charged-only particles at central rapidity.

\end{abstract}

\maketitle

\section{Introduction}
\label{sec:introduction}


For each high-energy nuclear collision, the magnetic field introduced by the original beam spectators will cause $P$ and $CP$ destruction of the left-handed particles in the QGP~\cite{PhysRevC.85.044907} via the electroweak interactions, and fluctuate event-by-event caused by fluctuation about proton positions~\cite{BZDAK2012171}, leading to an imbalance in the number of right- and left-handed$($anti$)$ particles. 
If the magnetic field was sufficiently strong, a significant charge separation along the direction of the magnetic field happens -- the CME~\cite{BLOCZYNSKI20131529}. 
To separate the CME-driven signal from the collective flow background especially second-order elliptic flow ($v_{2}$), one can measure the flow correlation $\gamma$ between charged hadron pairs of the same and opposite charges, then evaluate their difference $\Delta \gamma$.

Because the magnetic field was expected to be initially perpendicular to the reaction plane (RP) defined by beam axis and impact parameter, the "$\gamma$ correlation~\cite{PhysRevC.70.057901}" was defined as
\begin{equation}
\gamma_{\alpha \beta } =<cos(\phi_{\alpha}+\phi_{\beta}-2\Psi_{RP})>
\label{equ:gamma}
\end{equation} 
$\phi_{\alpha}$ and $\phi_{\beta}$ are the azimuthal angles of particles pairs of interest, and the averaging $<...>$ was performed over the pairs of particles and over events. 
Then the difference between the opposite-sign(OS) and same-sign(SS) $\gamma$ correlation called $\Delta \gamma$ was studied for the CME signal: 
\begin{equation} 
\Delta \gamma=\gamma_{OS}-\gamma_{SS}
\label{equ:delta_gamma}
\end{equation} 
%
since it was sensitive to the preferential emission of positively and negatively charged particles at the opposite sides of the RP.
In real experiments, $\Psi_{RP}$ was measured via either the spectator plane (SP) or participant plane(PP), defined by particles at most forward or central rapidity respectively.

Because the CME signal was usually weak relative to the collective flow background, the latter needs to be measured independently and accurately in relativistic heavy ion collisions, and various approaches have been applied to deal with the background~\cite{2006.05035,PhysRevC.97.044912}. The $\gamma$ correlation from background was defined usually as the equation below \cite{PhysRevC.70.057901,PhysRevC.81.064902,PhysRevC.81.031901,PhysRevC.83.014913,PhysRevC.101.034912}:

%
\begin{equation}
\Delta \gamma_{bkgd}=\frac{4N_{res}}{\langle N_{trk} \rangle ^{2}}<cos(\phi_{\alpha}-\phi_{\beta}-2\phi_{res})>v_{2,res}
\label{equ:delta_gamma_bkgd}
\end{equation}

%

%


The RHIC-STAR experiment proposed isobar run in 2018 with two collision systems: $_{44}^{96}Ru+_{44}^{96}Ru$ and $_{40}^{96}Zr+_{40}^{96}Zr$.
The strategy of two isobar species was to produce different signal but idential background,
because \ruru\ and \zrzr\ have the same number of nucleons but different numbers of protons.
Ideally, those charged+neutral quantities such as multiplicity and elliptic flow influence in $\Delta\gamma$ background, should be the same in the two systems. 
On the other hand, the difference of \ruru\ and \zrzr\ systems in the number of initial charges carried by protons within approximately equal volume can propagate to the initial magnetic fields generated, and measurable via the normalized charge separation correlation -- $\Delta\gamma/v_{2}$, removing those background due to resonance decay and local charge conservation. 

However, the STAR CME paper~\cite{PhysRevC.105.014901} showed the isobar ratio $(\ruru/\zrzr)$ of $\Delta\gamma/v_{2}$ below unity, seemingly opposite to the initial expectation of larger CME in \ruru\ for stronger magnetic field.
It was then important to cross-check the flow background especially $v_{2}$ for any possible bias, before applying their difference into the $\Delta\gamma/v_{2}$ observable. 
Since $v_2$ strongly dependent on the centrality and flow, and the measurements of latter both rely on their methods and particle selection,
it became necessary to study in details the effect of different electrical properties, (pseudo-)rapidity acceptance, and \pt\, thresholds on particles.

Therefore, we proposed the following analysis in AMPT for such test, since it was more convenient to expand and vary all these selections in the model, without constraints in real detectors performance.
The dependence found in AMPT, if any, could help removing possible bias in CME-signal searching at real experiments.
The next sections were organized as following: Sec.~\ref{sec:methods} explains the AMPT setup, Sec.~\ref{sec:results} lists the selection of particles generated by AMPT, and how the centrality and $v_2$ were reconstructed to compare with STAR results, then Sec.~\ref{sec:discussion} discusses how this analysis could help to expand the phase space in real experiments.

 


\section{Methods}
\label{sec:methods}

\subsection{AMPT, A multi-phase transport model}
We used AMPT~\cite{PhysRevC.72.064901,Nucl.Sci.Tech.32.113} model to generate about 2 million events for each \ruru\ and \zrzr\ collision systems. 
Owing to no magnetic field effect in AMPT model, it was expected to produce pure flow-background induced $\Delta\gamma\{\Psi_{SP}\}$ and $\Delta\gamma\{\Psi_{PP}\}$.
The AMPT model in its string-melting scenario was setup, including four stages: initial condition, partial scattering, hadronization, and hadronic interactions. 
(1)
The initial condition was the process in which two nuclei moving in opposite directions colliding and producing partons.
It determines which nucleons will collide based on the positions of the nuclei and the size of the scattering cross section.
These nucleons called participants generate partons and excitation strings,  
being simulated using Heavy-Ion Jet Interaction Generator (HIJING)~\cite{PhysRevD.44.3501,GYULASSY1994307}), giving the parton phase space distribution via excited strings and mini-jet partons mechanism.
(2)
Then at the process of partial scattering, the generated partons and excitation strings continue to move and collide thus evolving the parton phase space distribution. 
Such evolution equations were simulated by the ZPC~\cite{ZHANG1998193} model till freeze out, where the parton interactions include only two-body elastic scattering.
(3)
At the process of hadronization, the partons freeze into new hadrons through quark coalescence.
(4)
Finally, after hadronization, the hadrons undergo elastic and inelastic scattering again. AMPT simulates the inter-hadron transport processes by using the ART (A Relativistic Transport) model~\cite{JGXK2024090101,PhysRevC.52.2037}. 
Those nucleons previously excluded from scattering participants, called spectators on the other hand, also join this hadron cascade process.


The collision conditions of AMPT in this analysis were set up at \snn\ = 200 \GeV \, same as the STAR isobar runs \ruru\ and \zrzr\ systems, with initial nuclear thickness distributions based on Woods-Saxon parameters~\cite{PhysRevC.94.041901,PhysRevLett.127.242301,SHOU2015215},
\begin{equation}
\rho (r,\theta )=\frac{\rho _{0} }{1+exp[\frac{r-R_{0}(1+\beta _{2}Y_{2}^{0}(\theta )+\beta _{3}Y_{3}^{0}(\theta )) }{a} ]}
\label{equ:rho}
\end{equation}
where $r$ was radial distance from the center, $\theta$ was polar angle in spherical coordinates. 
Identical values for the radius $R_{0}$ and surface diffuseness $a$ were assumed for both Ru and Zr with no deformation present in either nucleus, while the $\rho_{0}$ also being same determined from the nucleon number A. 
The most significant parameters for describing the shape deformation of nuclei were characterized by the axial symmetric quadrupole deformation parameter $\beta_{2}$ and the octuple deformation parameter $\beta_{3}$.
 
\subsection{Impact parameter b distribution}
\begin{figure}[h!]
    \centering
    \includegraphics[width=0.8\linewidth]{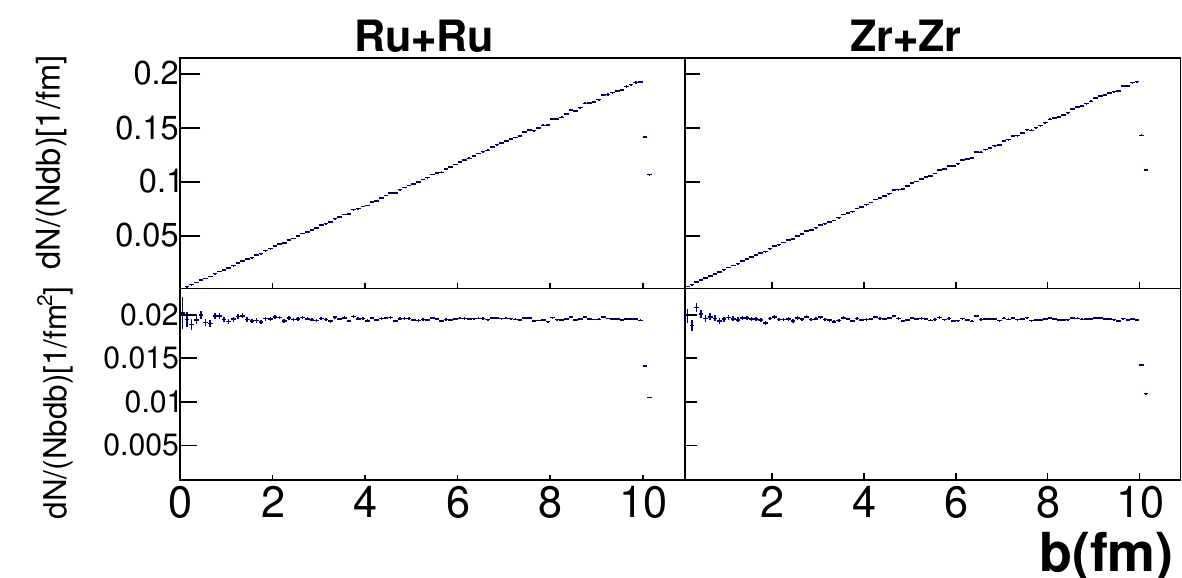}
    \caption{  
    The distributions of impact parameter b in the \ruru\ (left) and \zrzr\ (right) collision systems. All normalized to per event basis. 
    Upper: $\frac{{\rm d}N}{{\rm d}b} $ normalized to per event.
    Bottom: $\frac{{\rm d}N}{b{\rm d}b} $ normalized to per event.
    }
\label{fig:b-parameter}
\end{figure}

Because of the Lorentz-boost effect at STAR isobar energy, the relative distance of two colliding nuclei were expected to uniformly distributed in the transverse plane perpendicular to the beam axis.
This was verified by the flat distribution of $\frac{{\rm d}N}{b{\rm d}b}$ of impact parameter b at Fig.~\ref{fig:b-parameter},
from about 2 million AMPT events in each \ruru\ and \zrzr\ collision systems, setup from 0 to 10.2 fm according to the radius parameter $R$ defined by the Woods-Saxon distribution in Equ.~\ref{equ:rho}~\cite{PhysRevC.94.041901,XU2021136453}.
 
\subsection{Track selection}
%
In this analysis, we only retain the final-state hadrons as the equivalent of STAR tracks in their CME work~\cite{PhysRevC.105.014901}. 
One possible bias of STAR was that only charged particles at central rapidity were used to divide centrality intervals and measure $v_{2}$, mainly due to the constraints of detectors, e.g. the STAR TPC~\cite{ACKERMANN1999681} only detecting charged particles within specific pseudorapidity $|\eta|$ and \pt\, range.
This may be incomplete to the assumption of same flow background between \ruru\ and \zrzr\ based on the same number of inclusive final-state hadrons, both charged and neutral.
What was more, the $v_2$ was strongly dependent on centrality, which was also assumed from charged+neutral hadrons, but was measured at STAT CME work via the charged hadrons inside TPC.

On the other hand, the AMPT model can take full advantage of both the charged particles and neutral particles within a broad (pseudo-)rapidity and \pt\, coverage. 
This brings to our strategy in searching for possible bias in flow $v_{2}$ background to compare with the experimental results, by expanding the particle selection in both centrality and $v_2$ measurements. 

\begin{figure}[h!]
\centering
\includegraphics[width=0.8\linewidth]{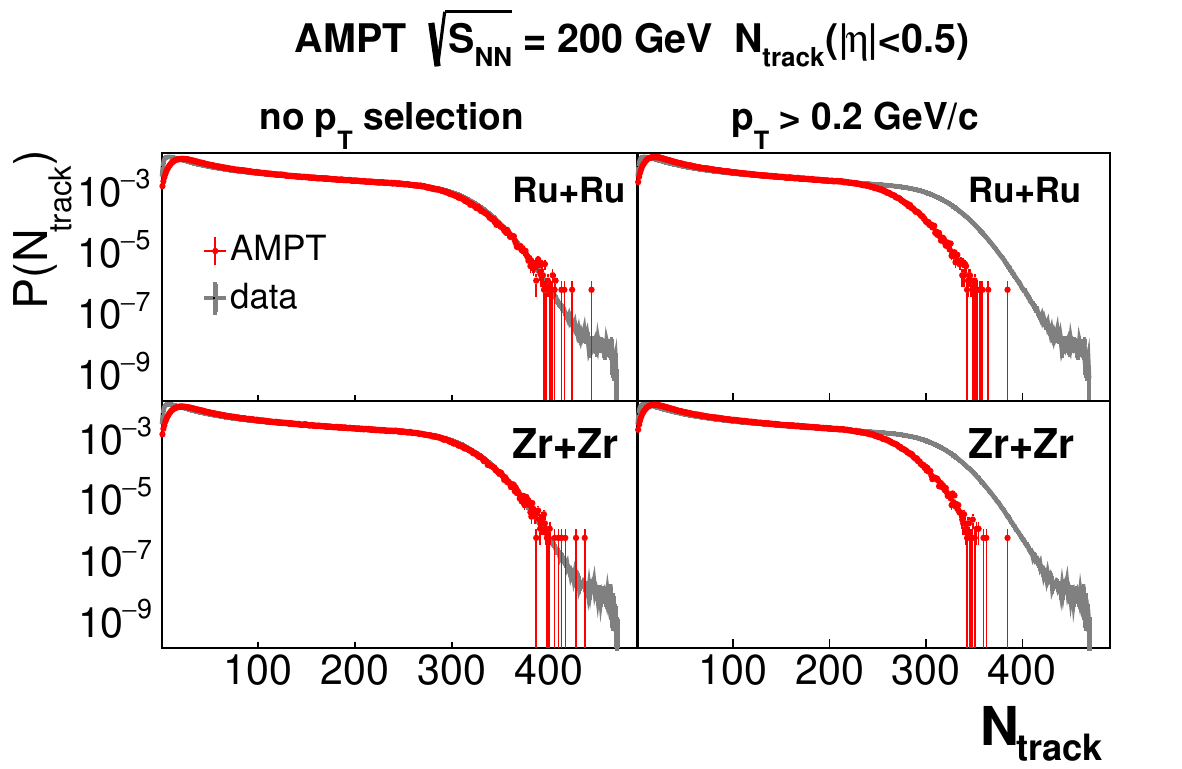}
\caption{
The charged particle distribution $N_{\mathrm{track} }$ from AMPT (red line) and experimental STAR~\cite{PhysRevC.105.014901} (gray line), at \snn\ = 200 \GeV\ in \ruru\ (upper panels) and \zrzr\ (bottom panels) collisions with acceptance $|\eta| < 0.5$, without (left side) and with (right side) \pt\ > 0.2 \GeV/$c$ selection. All were normalized by their own number of events.
}
\label{fig:Ntrack}
\end{figure}

\begin{figure}[h!]
\centering
\includegraphics[width=0.8\linewidth]{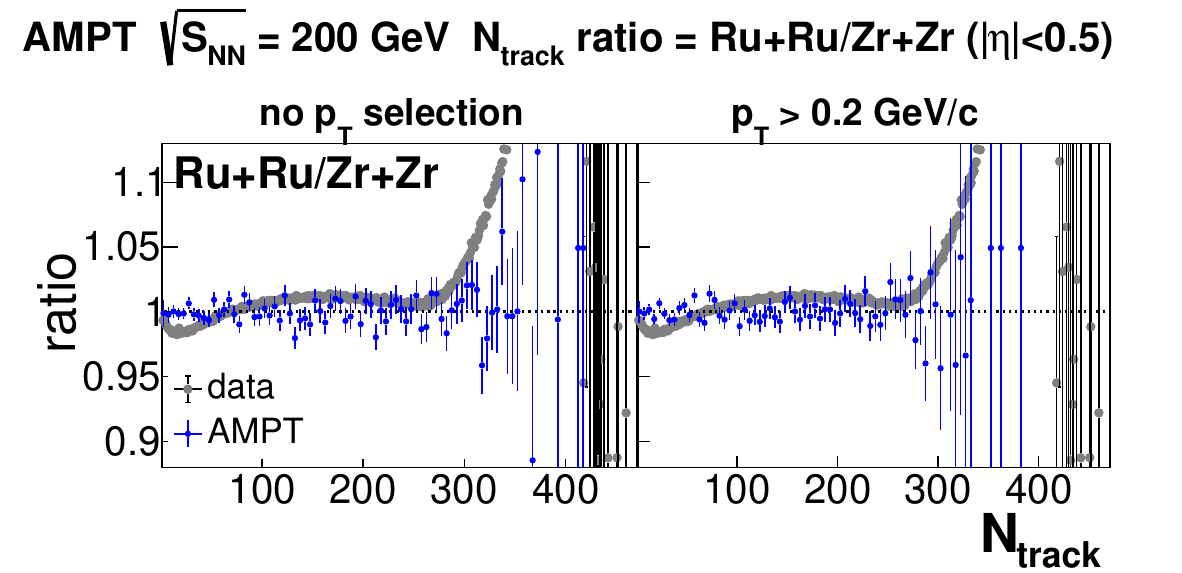}
\caption{ The \ruru\ to \zrzr\ ratio of $N_{\mathrm{track} }$ generated by AMPT (blue crosses) and STAR~\cite{PhysRevC.105.014901}) (gray line), at \snn\ = 200 \GeV\ collisions within acceptance $|\eta| < 0.5$, without (left side) and with (right side)  \pt $ > 0.2 $  \GeV/$c$\ selection.
%
%
}
\label{fig:Ntrack_ratio}
\end{figure}

First, as a sanity check, we test with only charged particles within $|\eta| < 0.5$ for centrality determination, the same as STAR~\cite{PhysRevC.105.014901}.
Fig.~\ref{fig:Ntrack} shows $P_{{\rm trk}}$, the probability distribution of $N_{trk}$ 
generated by AMPT model in \ruru\ and \zrzr\ collision systems, also comparing with the luminosity and $V_{Z,TPC}$ corrected distributions $P(N_{{\rm trk}}^{{\rm offline}})$ by STAR.
Then Fig.~\ref{fig:Ntrack_ratio} compared their ratio of \ruru\ to \zrzr\ collision system 
between AMPT model with STAR results.
Both figures show that those AMPT work without \pt\ limitation reproduce the STAR experimental results better.
Note that AMPT model does not include detector effects such as efficiency or ghost particles.

%

This qualitative agreement brought us confidence to make multiple tests to look for possible bias in determining the flow background with specific particle selection.
The selected particles would vary 
without or with requiring \pt\ $ > 0.2$ GeV/$c$ as Fig.~\ref{fig:Ntrack} and \ref{fig:Ntrack_ratio} indicated,
in addition to the limit of pseudorapidity acceptance $|\eta| < 0.5$.
For each set of particle selection, both centrality and flow would be recalculated at AMPT.
We expect that such charged+neutral comparisons will indicate the effect of possible bias .

\section{Results}
\label{sec:results}

\subsection{Centrality determination}
%
Centrality was intended to describe the degree of deviation of the centers of the two nuclei in a heavy ion collision. 
Theoretically, the centrality of a collision event was defined by the percentile of the total cross section, usually in hard scattering sense. 
\begin{equation}
\mathrm{Centrality}(b_1)=\frac{\sigma_{b_{1} }}{\sigma_{Total}  } =\frac{\int_{o}^{b_{1} }\frac{{\rm d}\sigma}{{\rm d} b}db }{\int_{0}^{\infty} \frac{{\rm d}\sigma}{{\rm d} b}db }
\label{equ:centrality-intelgral-b}
\end{equation}
 %
In the real experiments, however, the collision impact parameter $b$ was not directly measurable, 
but determined statistically through the correlation of following:
the smaller the collision parameter, the bigger the overlap region of the two nuclei, the more participants in the collision, and the more final-state particles were generated. 
The reported centrality was then measured based on the final-state particles production into average percentage bins.

One approach was called the integration method, where based on multiplicity distribution in the horizontal axis of Fig.~\ref{fig:Ntrack} like from either model or experimental data, the integral edge cuts were made from right to left to define the upper and lower limits of $N_{trk}$ for each centrality interval based on the ascending percentile of the integral area.

\begin{figure}[h!]
\centering
\includegraphics[width=0.8\linewidth, inner]{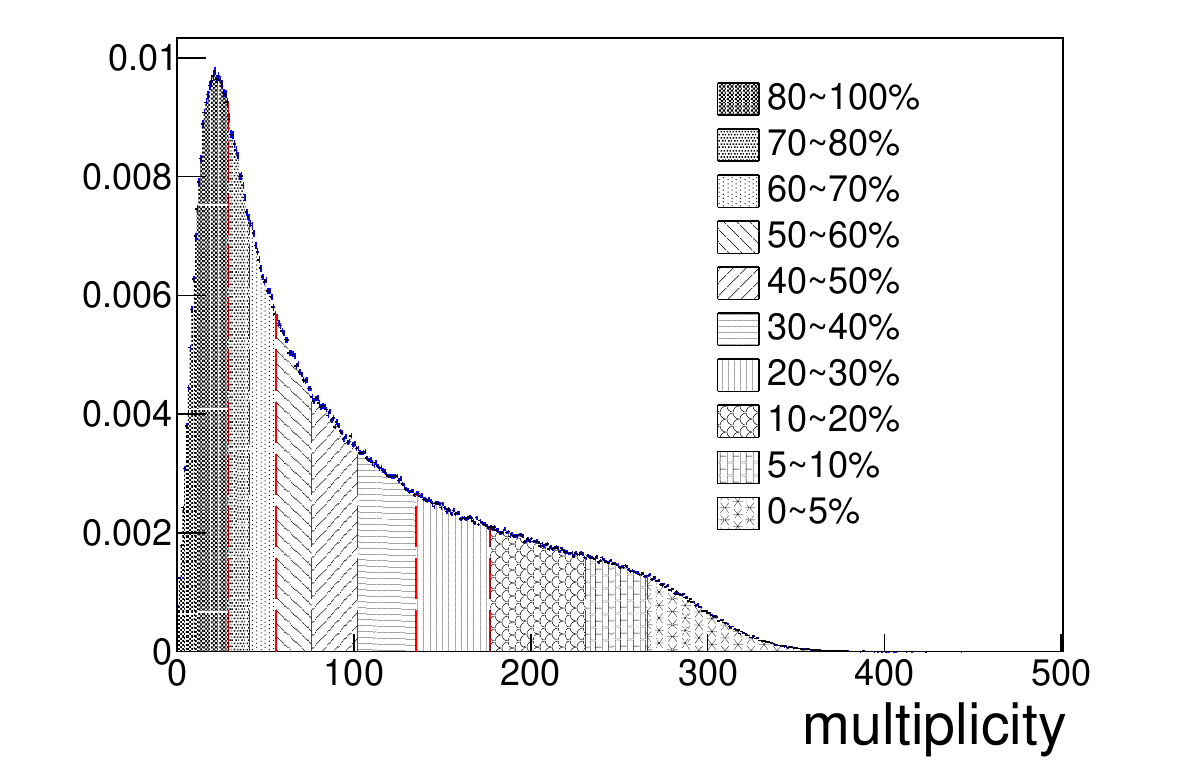}
\includegraphics[width=0.8\linewidth, inner]{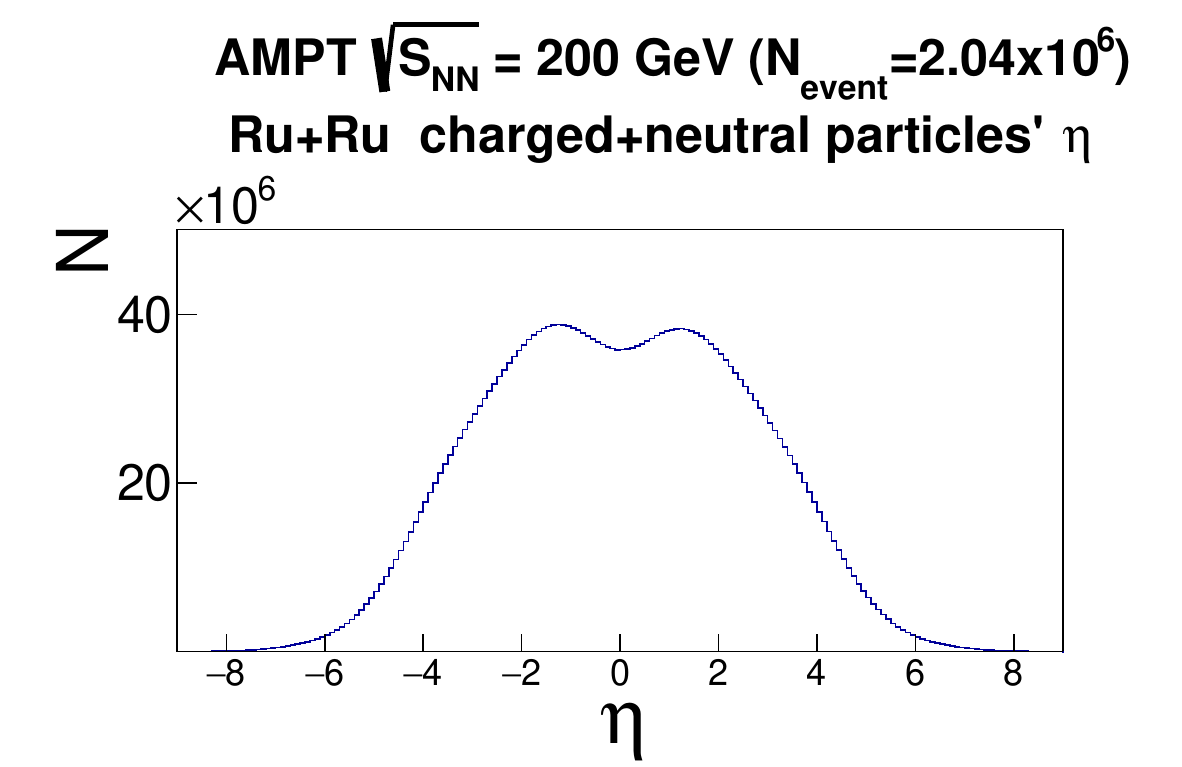}
\caption{One sample of integral centrality method. The top panel presents the multiplicity distribution from AMPT within $|\eta|<0.5$ and without \pt\ selection in \ruru\ system. The bottom panel presents how broad (pseudo-)rapidity range the AMPT generated particles extend to.}
\label{fig:centrality_define}
\end{figure}

This method was evaluated with our AMPT work, with all centrality intervals at AMPT were grouped in the same percentile as STAR did in Tab.~\ref{tab:1},
but with multiple particle categories (charged and/or neutral) and (pseudo-)rapidity acceptance selection in \ruru\ or \zrzr\ collision systems.
One sample among all these selections was shown at Fig.~\ref{fig:centrality_define}. 
The top panel shows the integration method for \ruru, whose particle count came from a subset of the bottom panel, where the particles generated by AMPT extend to a much broader (pseudo-)rapidity coverage than the $\eta $ acceptance of STAR TPC.
This supported our approach to apply a relative broader acceptance of particle selection at AMPT than STAR in later sections.

Another approach was called the designation method, where the upper and lower limits of each centrality interval at AMPT were calculated by multiplying charge-dependent coefficients to corresponding STAR numbers~\cite{PhysRevC.105.014901} at Tab.~\ref{tab:1}, 
which were derived via integration + MC Glauber method.
These coefficient were necessary because the STAR CME work chose only charged particles as track, while AMPT results could select both charged and/or neutral particles for $N_{trk}$, in addition to variant $\eta$ and \pt \, selections.

The two methods were combined as cross check.
First, the integration method was applied for each combination of particle selection to decide the centrality intervals by AMPT itself.
Samples of this integration methods, using charged-only particles at same $\eta$ and \pt\, thresholds as was done in STAR, were shown in Fig.~\ref{fig:Ntrack} and \ref{fig:Ntrack_ratio} indicated,
and numbers listed in Tab.~\ref{tab:3} and \ref{tab:4} later.

Then, for the designation method of this particle selection, the coefficients of charged+neutral to charged were either the naive expectation of 1.5 based on fragmentation without isospin or nuclear effect,
or calculated by comparing the upper and lower limit numbers of charged and/or neutral particles generated by AMPT as Tab.~\ref{tab:2} summarized,
approximately 1.73 for charged+neutral hadrons, and 1.65 for pions, the highest proportion in hadrons, by calculating $\pi^{+}$+$\pi^{-}+\pi^{0}$ to $\pi^{+}$+$\pi^{-}$.
It also drew our attention that the difference of coefficients between \ruru\ and \zrzr\ was smaller than that of initial ratio of charge to atom number between the two collision systems, i.e. 44/96 to 40/96.
If AMPT used only neutral particles for centrality, then the coefficients above were all subtracted by one.
And the coefficients would be naturally one if AMPT used only charged particles.

\begin{table}[h!]
      \caption{Centrality definition by STAR experiment using integration + MC Glauber model method. }
      \centering
      \begin{tabular}{@{}ccccc}
      \hline
           &  \multicolumn{2}{c}{\ruru}&  \multicolumn{2}{c}{\zrzr}\\
      \hline
           Centrality$(\%)$ & $ N_{trk} $ &  $\langle N_{trk} \rangle$ &  $ N_{trk} $ & $\langle N_{trk} \rangle$ \\
           \hline
            0-5 &  258-500&  289.32&  256-500& 287.36\\
            5-10&  216-257&  236.30&  213-255& 233.79\\
           10-20&  151-215&  181.76&  147-212& 178.19\\
           20-30&  103-150&  125.84&  100-146& 122.35\\
           30-40&   69-102&   85.22&   65-99 &  81.62\\
           40-50&   44-68 &   55.91&   41-64 &  52.41\\
           50-60&   26-43 &   34.58&   25-40 &  32.66\\
           60-70&   15-25 &   20.34&   14-24 &  19.34\\
           70-80&    8-14 &   11.47&    7-13 &  10.48\\
 \hline
      \end{tabular}
      \label{tab:1}
  \end{table}
  
\begin{table}[h]
      \caption{The ratio of charged + neutral particles to charged particles, namely (c+n)/c, and  $\pi^{+}$+$\pi^{-}$ to $\pi^{0}$  at different $\eta$ and \pt\ selection by AMPT model.}
      \centering
      \begin{tabular}{ccllcclcl}
      \hline
 & \multicolumn{4}{c}{\ruru} \\ 
 particles selected& \multicolumn{2}{c}{ $\left | \eta \right | <0.5$}& \multicolumn{2}{c}{no $\eta$ cut} \\ 
            ratio& no \pt\ cut & \pt\ > 0.2 GeV/$c$ &  no \pt\ cut & \pt\ > 0.2 GeV/$c$ \\ 
  \hline
            (c+n)/c&   1.76&1.73&   1.75&1.71 \\ 
            ($\pi^{+}$+$\pi^{-}$)/$\pi^{0}$)&   1.69&1.65&   1.68&1.62 \\ 
            \hline

             & \multicolumn{4}{c}{\zrzr}\\
 particles selected& \multicolumn{2}{c}{ $\left | \eta \right | <0.5$}& \multicolumn{2}{c}{no $\eta$ cut}\\
            ratio& no \pt\ cut & \pt\ > 0.2 GeV/$c$ &  no \pt\ cut & \pt\ > 0.2 GeV/$c$ \\
  \hline
          (c+n)/c&      1.76&1.73&  1.75&1.71\\
            ($\pi^{+}$+$\pi^{-}$)/$\pi^{0}$)&   1.69&1.65&  1.68&1.62\\
            \hline

      \end{tabular}
      \label{tab:2}
  \end{table}

\subsection{The mean multiplicity}
For direct comparison with STAR at similar $N_{trk}$ distributions, we first test with integration method in the same way STAR did, by selecting charged-only particles within $|\eta| < 0.5$ in \ruru\ and \zrzr\ collisions at \snn\ = 200 \GeV.
Both $N_{trk}$ distributions without and with \pt\ > 0.2 GeV/$c$ selection were shown in Fig.~\ref{fig:Ntrack} and ~\ref{fig:Ntrack_ratio},
and the corresponding numbers were listed in Tab~\ref{tab:3} and~\ref{tab:4}, respectively without and with \pt\, selection.
The mean multiplicity $\langle N_{trk} \rangle$ was also calculated in corresponding centrality intervals.
The relative difference of $\langle N_{trk} \rangle$ between \ruru\ and \zrzr\ was found to be smaller than that of charge to atomic number between the two collision systems.
This was consistent with what was found previously with the designation method at Tab.~\ref{tab:2}.

On the other hand, even if the AMPT parameters of $a$ and $R_{0}$ and deformation were setup according to STAR \ruru\ and \zrzr\ collision systems,
and similar $\eta $ and \pt\ selections applied to particles of STAR results,
Fig.~\ref{fig:Ntrack} and \ref{fig:Ntrack_ratio} have shown deviation of the per-event normalized $N_{trk}$ distribution and their ratios between model and real data,
with details listed in Tab.~\ref{tab:1}, \ref{tab:3}, and \ref{tab:4}.
This was possibly due to the nuclear deformation mechanism~\cite{PhysRevLett.128.022301} where Zr holds an octopole deformation and Ru being ellipsoidal,
and it requires more careful handling at AMPT especially for most central or most peripheral collisions.

As a result, because for the designation method the definition of centrality and later flow background requires more strict correspondence of the mean multiplicity $\langle N_{trk} \rangle$ between AMPT and STAR,
we felt the integration method was more feasible to our purpose,
since the latter reflects the actual percentile regions of central and peripheral in AMPT, and more flexible to adjust the $\langle N_{trk} \rangle$ for centrality and $v_2$ background.
The designation method would mainly provide a reference in the later sections.

\begin{table}[h]
    \caption{AMPT model centrality definition by $N_{trk}$ ranges (charged track multiplicity within $|\eta| < 0.5$ and without \pt\ selection) in \ruru\ and \zrzr\ collisions at \snn = 200 \GeV. The centrality intervals were grouped to the same percentiles as STAR CME work~\cite{PhysRevC.105.014901}, where $N_{trk}$ was the multiplicity range, and $\langle N_{trk} \rangle$ was the mean multiplicity within.}
    \centering
    \begin{tabular}{c|cr|cr}
    \hline
    \hline
 \snn = 200 \GeV & \multicolumn{2}{c}{\ruru}& \multicolumn{2}{c}{\zrzr}\\
 \hline
    Centrality$(\%)$ & $ N_{trk} $ & $\langle N_{trk} \rangle$ & $ N_{trk} $ & $\langle N_{trk} \rangle$ \\
     \hline
       0-5 & 265-500 & 292.44 & 265-500 & 291.74\\ 
      5-10 & 231-264 & 247.31 & 231-264 & 246.92\\
     10-20 & 177-230 & 202.28 & 177-230 & 202.31\\
     20-30 & 135-176 & 154.68 & 135-176 & 154.62\\
     30-40 & 102-134 & 117.31 & 102-134 & 117.34\\ 
     40-50 &  76-101 &  87.97 &  76-101 &  87.97\\  
     50-60 &  56-75  &  65.07 &  56-75  &  65.08\\ 
     60-70 &  41-55  &  47.68 &  41-55  &  47.68\\  
     70-80 &  29-40  &  34.28 &  29-40  &  34.29\\ 
     80-100 &  1-28  &  17.28 &   1-28  &  17.27\\  
      \hline
      \hline
    \end{tabular}
    \label{tab:3}
\end{table}

\begin{table}[h]
    \caption{AMPT model centrality definition by $N_{trk}$ ranges (charged track multiplicity within $|\eta| < 0.5$ and \pt\ > 0.2 GeV/$c$) in \ruru\ and \zrzr\ collisions at \snn = 200 \GeV. The centrality intervals were grouped to the same percentiles as STAR CME work~\cite{PhysRevC.105.014901}, where $N_{trk}$ was the multiplicity range, and $\langle N_{trk} \rangle$ was the mean multiplicity within.}
    \centering
    \begin{tabular}{c|cr|cr}
    \hline
    \hline
 \snn = 200 \GeV & \multicolumn{2}{c}{\ruru}& \multicolumn{2}{c}{\zrzr}\\
 \hline
    Centrality$(\%)$ & $ N_{trk} $ & $\langle N_{trk} \rangle $ & $ N_{trk} $ & $\langle N_{trk} \rangle$ \\
     \hline
      0-5   & 230-500 & 253.31 & 230-500 & 253.22\\ 
      5-10  & 201-229 & 214.44 & 201-229 & 214.45\\
     10-20  & 154-200 & 175.95 & 154-200 & 175.96\\
     20-30  & 117-153 & 134.27 & 117-153 & 134.21\\
     30-40  &  89-116 & 101.96 &  89-116 & 101.96\\ 
     40-50  &  66-88  &  76.53 &  66-88  &  76.54\\  
     50-60  &  49-65  &  56.64 &  49-65  &  56.66\\ 
     60-70  &  35-48  &  41.19 &  35-48  &  41.17\\  
     70-80  &  25-34  &  29.33 &  25-34  &  29.33\\ 
     80-100 &   1-24  &  14.78 &   1-24  &  14.76\\  
      \hline
      \hline
    \end{tabular}
    \label{tab:4}
\end{table}

\begin{figure}[h!]
\centering
\includegraphics[width=0.8\linewidth]{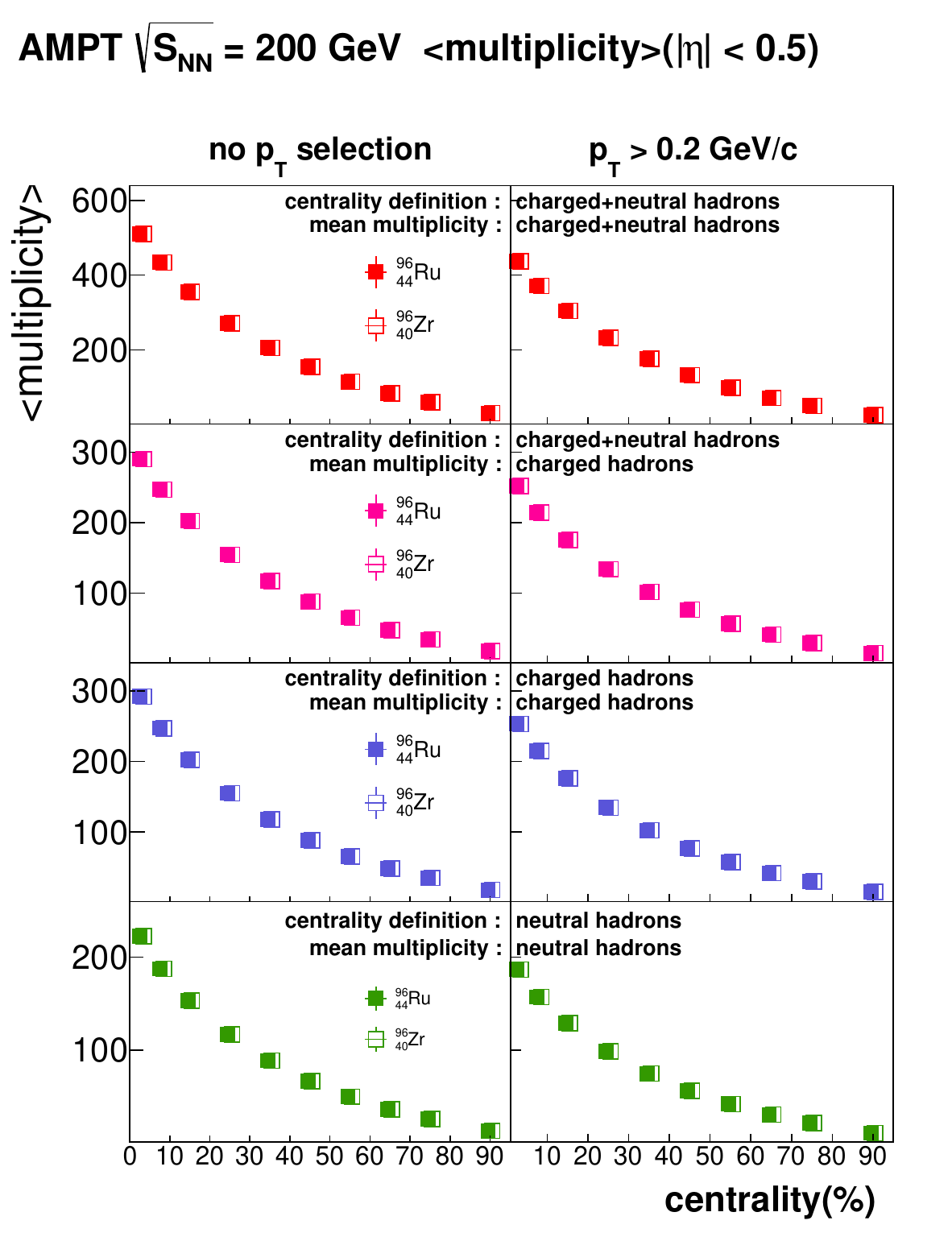}
\caption{
The mean multiplicity distributions with centrality intervals at \ruru\ (solid) and \zrzr\ (open) from AMPT, without (left) or with (right) \pt\ > 0.2 GeV/$c$ selection. Both multiplicity and centrality were calculated using multiple particle categories of charged and/or neutral ones.
For direct comparison with the experimental ratio by STAR, these AMPT results also applied $|\eta| < 0.5 $ for particles. 
}
\label{fig:meanmultiplicity_etacut0.5}
\end{figure}
\begin{figure}[h!]
\centering
\includegraphics[width=0.98\linewidth]{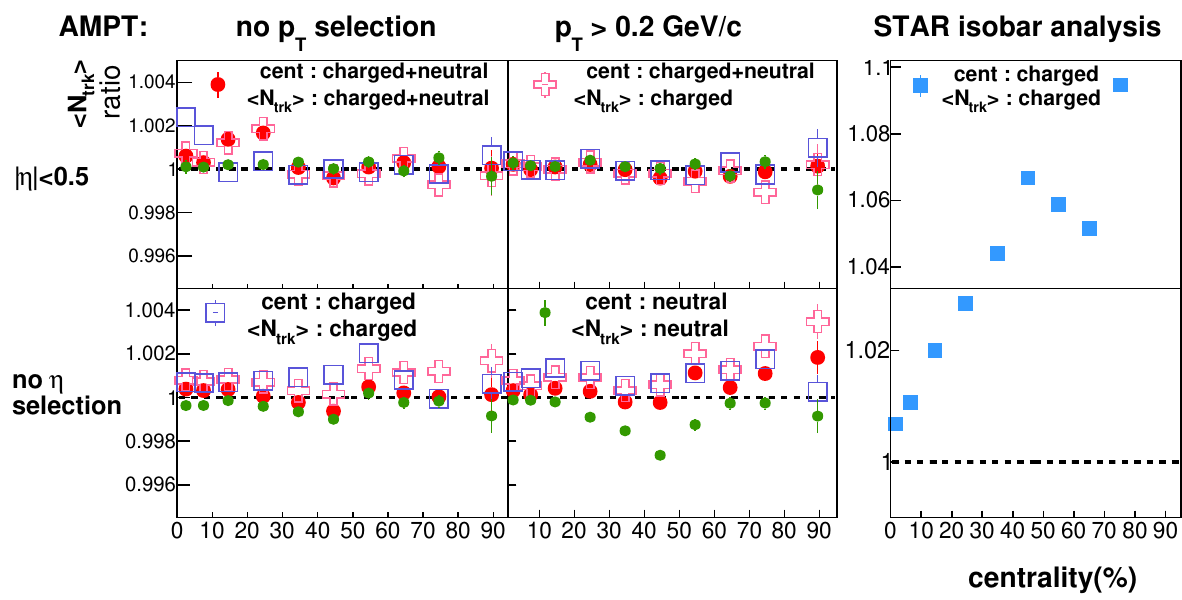}
\caption{
In the left four panels, we presented the ratios of mean multiplicity distributions of \ruru\ to \zrzr\ by AMPT.
Both centrality and mean multiplicity were calculated using multiple particle categories of charged and/or neutral ones.
They were presented without (left) or with (right) \pt\ > 0.2 GeV/$c$ selection, and with $|\eta| < 0.5 $ (upper) or without (bottom) $\eta$ selection.
In the most right panel, the STAR data~\cite{PhysRevC.105.014901} using only charged particles for centrality and multiplicity was shown as blue solid points for comparison.
}
\label{fig:meanmultiplicity_ratio}
\end{figure}

Next, it was important to test as many combinations of charged and/or neutral particles as possible, in the calculation of either centrality interval or mean multiplicity.
Fig.~\ref{fig:meanmultiplicity_etacut0.5} showed the $\langle N_{trk} \rangle$ in \ruru\ and \zrzr\ as a function of centrality, 
then comparing their ratios with STAR data in the upper panels of Fig.~\ref{fig:meanmultiplicity_ratio},
both without or with  \pt\ > 0.2 GeV/$c$selection.
No significant difference between \ruru\ and \zrzr\ was observed for either charge property, while we had expected more charged protons in \ruru\ would generate more charged tracks than \zrzr\ in AMPT like what happened at STAR.

\begin{figure}[h!]
\centering
\includegraphics[width=0.8\linewidth]{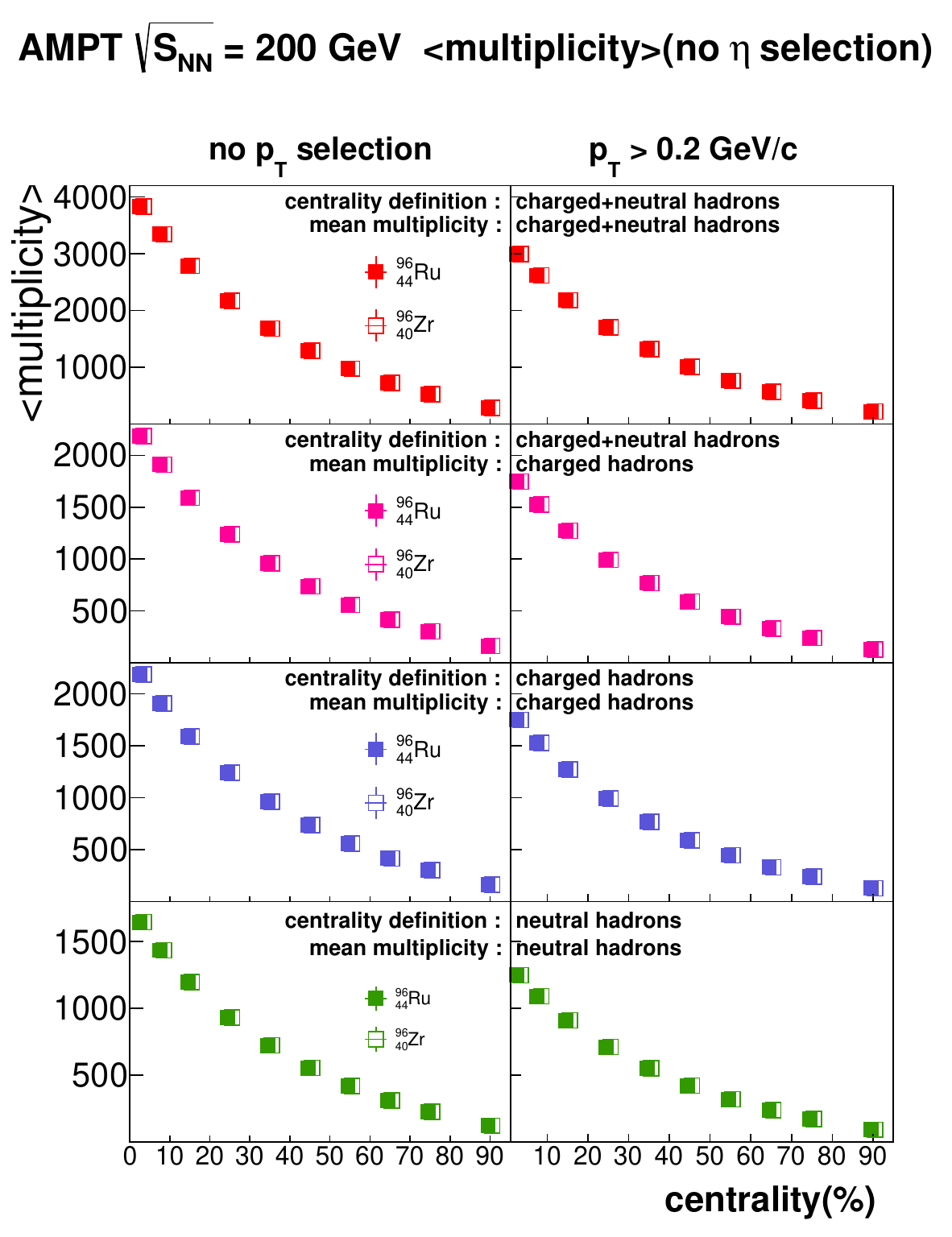}
\caption{
The mean multiplicity distributions with centrality intervals at \ruru\ (solid) and \zrzr\ (open) from AMPT, without (left) or with (right) \pt\ > 0.2 GeV/$c$ selection. 
Both multiplicity and centrality were calculated using multiple particle categories of charged and/or neutral ones.
For complacent comparison with the experimental ratio by STAR, these AMPT results applied no $\eta $ selection for particles. 
}
\label{fig:meanmultiplicity_noetacut}
\end{figure}

The deviation with expectation above was possibly caused by the broad $\eta$ distribution of AMPT particles in Fig.~\ref{fig:centrality_define},
being unable to reproduce $N_{trk}$ at STAR as was shown at Fig.~\ref{fig:Ntrack} at same $\eta$ limit.
It drove new tests with $\eta$ limit released in AMPT particle selection,
with the $\langle N_{trk} \rangle$ in \ruru\ and \zrzr\ as a function of centrality shown in Fig.~\ref{fig:meanmultiplicity_noetacut},
then their ratios compared with STAR data in the lower two panels of Fig.~\ref{fig:meanmultiplicity_ratio},
both without or with  \pt\ > 0.2 GeV/$c$ selection.
While the $N_{trk}$ distributions of AMPT was enhanced then over STAR, the ratios of $\langle N_{trk} \rangle$ between \ruru\ and \zrzr\ at AMPT in Fig.~\ref{fig:meanmultiplicity_ratio} mostly still stayed around unity within one percent level, unlike the STAR results beside.

On the other hand, now that we released the $\eta$ limit at the bottom two panels of Fig.~\ref{fig:meanmultiplicity_ratio},
those ratios of $\langle N_{trk} \rangle$ using only charged particles
slightly increased from unity with centrality and deviate those ratios of charged+neutral particles.
Such trend indicated if all partons outputs by AMPT were retained, the initial charge asymmetry were slightly better conserved at the hadron stage later, 
thus shifting the boundaries of centrality intervals based on charged-only multiplicity. 
This seemed to be supported by the ratios of neutral-only $\langle N_{trk} \rangle$ shifting from unity in the opposite direction at the bottom left panel of Fig.~\ref{fig:meanmultiplicity_ratio}, also with $\eta$ limit released.

Such shift from unity at AMPT, either upward or downward, indicated a possible bias in flow background by real experimental data when only charged particles were selected.
This was because ideally the charged+neutral $\langle N_{trk} \rangle$ should be chosen when calculating centrality and later flow background,
and their ratio between \ruru\ and \zrzr\ had been expected to stay with unity, 
since the two isobar systems only differ in charged protons but hold same nucleons.
This was exactly what we observed in Fig.~\ref{fig:meanmultiplicity_ratio} with charge+neutral particles selected.
The actual numbers of centrality definition by $N_{trk}$ and $\langle N_{trk} \rangle$ generated by AMPT model, with all combinations of electrical property, $\eta$, and \pt, were listed in Table~\ref{tab:5}, ~\ref{tab:6}, ~\ref{tab:7}, ~\ref{tab:8}.

\begin{table}
    \caption{
    Centrality definition at AMPT model \ruru\ at \snn\ = 200 \GeV, by $N_{trk}$ ranges without any $\eta$ or \pt\ selection, in categories of charge+neutral particles, charged only, and neutral only. For each definition of centrality, the $\langle N_{trk} \rangle $ were also calculated by charged and/or neutral particles in each centrality interval.   
    }
    \centering
    \begin{tabular}{c|crrcrcr|}
    \hline
    \hline
 &     &&\multicolumn{2}{c}{\ruru}&&\multicolumn{2}{c}{}\\
 \hline
    multiplicity& \multicolumn{3}{c}{charged+neutral particles}&\multicolumn{2}{c}{charged particles}&\multicolumn{2}{c}{neutral particles}\\
 Centrality$(\%)$ & $ N_{trk} $ & $\langle N_{trk} \rangle_{c+n} $& $\langle N_{trk} \rangle_c $ & $ N_{trk} $ & $\langle N_{trk} \rangle_c $& $ N_{trk} $ & $\langle N_{trk} \rangle_n $\\
     \hline
       0-5 & 3555-6000& 3827.37& 2185.91& 2031-3000& 2187.91& 1525-2100& 1644.25\\ 
      5-10 & 3143-3554& 3344.06& 1910.19& 1795-2030& 1909.96& 1349-1524& 1434.30\\
     10-20 & 2456-3142& 2785.01& 1590.69& 1403-1794& 1590.53& 1054-1348& 1194.87\\
     20-30 & 1910-2455& 2172.39& 1240.81& 1091-1402& 1240.72&  820-1053&  932.10\\
     30-40 & 1472-1909& 1682.6& 961.01&  842-1090& 961.35&  632-819&  722.01\\ 
     40-50 & 1123-1471& 1290.55&  737.28&  642-841&  737.82&  482-631&  553.60\\  
     50-60 &  843-1122& 977.09&  558.19&  482-641&  558.52&  362-481&  419.27\\ 
     60-70 &  621-842&  726.99&  415.44&  355-481&  415.44&  267-361&  312.07\\  
     70-80 &  446-620&  529.787&  302.75&  255-354&  302.61&  192-266&  227.61\\ 
     80-100&    1-445&  285.31&  163.01&    1-254&  162.68&    1-191&  122.23\\  
      \hline
      \hline
    \end{tabular}
    \label{tab:5}
\end{table}

\begin{table}
    \caption{
    Centrality definition at AMPT model \zrzr\ at \snn\ = 200 \GeV, by $N_{trk}$ ranges without any $\eta$ or \pt\ selection, in categories of charge+neutral particles, charged only, and neutral only. For each definition of centrality, the $\langle N_{trk} \rangle $ were also calculated by charged and/or neutral particles in each centrality interval.   
    }    \centering
    \begin{tabular}{c|crrcrcr|}
    \hline
    \hline
 &     &&\multicolumn{2}{c}{\zrzr}&&\multicolumn{2}{c}{}\\
 \hline
   multiplicity& \multicolumn{3}{c}{charged+neutral }&\multicolumn{2}{c}{charged }&\multicolumn{2}{c}{neutral}\\
 Centrality$(\%)$ & $ N_{trk} $ & $\langle N_{trk} \rangle_{c+n} $& $\langle N_{trk} \rangle_c $ & $ N_{trk} $ & $\langle N_{trk} \rangle_c $& $ N_{trk} $ & $\langle N_{trk} \rangle_n $\\
      \hline
      0-5  & 3554-6000& 3825.90& 2184.17& 2030-3000& 2186.38& 1526-2100& 1644.86\\ 
      5-10 & 3142-3553& 3343.00& 1908.79& 1794-2029& 1908.69& 1349-1525& 1434.82\\
     10-20 & 2455-3141& 2783.98& 1589.37& 1402-1793& 1589.42& 1054-1348& 1195.03\\
     20-30 & 1910-2454& 2172.26& 1239.93& 1090-1401& 1239.77&  820-1053&  932.48\\
     30-40 & 1473-1909& 1682.98& 960.71&  841-1089& 960.45&  633-819&  722.49\\ 
     40-50 & 1123-1472& 1291.36&  737.17&  641-840&  737.05&  482-632&  554.15\\  
     50-60 &  842-1122& 976.62&  557.45&  481-640&  557.40&  362-481&  419.19\\ 
     60-70 &  621-841&  726.84&  414.96&  355-480&  415.11&  267-361&  312.14\\  
     70-80 &  446-620&  529.77&  302.38&  255-354&  302.63&  192-266&  227.65\\ 
     80-100&    1-445&  285.27&  162.74&    1-254&  162.58&    1-191&  122.33\\  
      \hline
      \hline
    \end{tabular}
    \label{tab:6}
\end{table}

\begin{table}
    \caption{
    Centrality definition at AMPT model \ruru\ at \snn\ = 200 \GeV, by $N_{trk}$ ranges without any $\eta$ selection and \pt\ > 0.2 GeV/$c$, in categories of charge+neutral particles, charged only, and neutral only. For each definition of centrality, the $\langle N_{trk} \rangle $ were also calculated by charged and/or neutral particles in each centrality interval. 
    }
    \centering
    \begin{tabular}{c|crrcrcr|}
    \hline
    \hline
 &     &&\multicolumn{2}{c}{\ruru}&&\multicolumn{2}{c}{}\\
 \hline
    multiplicity& \multicolumn{3}{c}{charged+neutral particles}&\multicolumn{2}{c}{charged particles}&\multicolumn{2}{c}{neutral particles}\\
 Centrality$(\%)$ & $ N_{trk} $ & $\langle N_{trk} \rangle_{c+n} $& $\langle N_{trk} \rangle_c $ & $ N_{trk} $ & $\langle N_{trk} \rangle_c $& $ N_{trk} $ & $\langle N_{trk} \rangle_n $\\
     \hline
      0-5  & 2780-6000 & 2993.61 & 1747.87 & 1624-3000 & 1749.92 & 1159-3000 &1248.51\\ 
      5-10 & 2461-2779 & 2617.19 & 1527.47 & 1437-1623 & 1527.90 & 1026-1158 &1090.34\\
     10-20 & 1925-2460 & 2181.96 & 1272.94 & 1123-1436 & 1273.34 &  803-1025 & 909.51\\
     20-30 & 1498-1924 & 1703.40 &  993.33 &  874-1122 &  993.64 &  625-802  & 710.28\\
     30-40 & 1154-1497 & 1319.40 &  769.08 &   673-873 &  769.33 &  481-624  & 550.43\\ 
     40-50 &  880-1153 & 1011.35 &  589.52 &   513-672 &  589.61 &  367-480  & 421.73\\  
     50-60 &  659-879  &  764.75 &  445.63 &   384-512 &  445.64 &  275-366  & 318.92\\ 
     60-70 &  484-658  &  567.48 &  330.63 &   283-383 &  330.63 &  203-274  & 237.07\\  
     70-80 &  347-483  &  412.42 &  240.25 &   202-282 &  240.25 &  145-202  & 172.40\\ 
     80-100&    1-346  &  221.28 &  128.76 &     1-202 &  128.76 &    1-144  &  92.14\\  
      \hline
      \hline
    \end{tabular}
    \label{tab:7}
\end{table}

\begin{table}
    \caption{
    Centrality definition at AMPT model \zrzr\ at \snn\ = 200 \GeV, by $N_{trk}$ ranges without any $\eta$ selection and \pt\ > 0.2 GeV/$c$, in categories of charge+neutral particles, charged only, and neutral only. 
    For each definition of centrality, the $\langle N_{trk} \rangle $ were also calculated by charged and/or neutral particles in each centrality interval. 
    }
    \centering
    \begin{tabular}{c|crrcrcr|}
    \hline
    \hline
 &     &&\multicolumn{2}{c}{\zrzr}&&\multicolumn{2}{c}{}\\
 \hline
    multiplicity& \multicolumn{3}{c}{charged+neutral particles}&\multicolumn{2}{c}{charged particles}&\multicolumn{2}{c}{neutral particles}\\
 Centrality$(\%)$ & $N_{trk}$ & $\langle N_{trk} \rangle_{c+n}$ & $\langle N_{trk} \rangle_c$ & $N_{trk}$ & $\langle N_{trk} \rangle_c$ & $N_{trk}$ & $\langle N_{trk} \rangle_n$\\
     \hline
     0-5   & 2781-6000 & 2992.67 & 1746.51 & 1623-3000 & 1748.35 & 1159-3000 & 1248.65\\ 
     5-10  & 2461-2780 & 2616.86 & 1526.66 & 1435-1622 & 1526.54 & 1026-1158 & 1090.48\\
     10-20 & 1924-2460 & 2181.04 & 1271.76 & 1122-1434 & 1271.62 &  803-1025 &  909.69\\
     20-30 & 1498-1923 & 1702.92 &  992.42 &  873-1121 &  992.42 &  626-802  &  710.92\\
     30-40 & 1155-1497 & 1319.68 &  768.83 &  673-872  &  768.91 &  483-625  &  551.27\\ 
     40-50 &  879-1154 & 1011.58 &  589.17 &  512-672  &  589.23 &  368-482  &  422.85\\  
     50-60 &  658-878  &  763.88 &  444.74 &  384-511  &  445.07 &  275-367  &  319.22\\ 
     60-70 &  484-657  &  567.22 &  330.21 &  282-383  &  330.54 &  203-274  &  237.13\\  
     70-80 &  346-483  &  411.97 &  239.69 &  202-281  &  240.02 &  145-202  &  172.45\\ 
     80-100&    1-345  &  210.88 &  128.31 &    1-201  &  128.33 &    1-144  &   92.22\\  
      \hline
      \hline
    \end{tabular}
    \label{tab:8}
\end{table}

\subsection{flow measurement}

In heavy-ion collisions, anisotropic flow was created as a response to the anisotropies in the initial geometry and the evolution of this system of hot and dense matter.
These flow were characterized by coefficients in the Fourier expansion of the azimuthal dependence of the invariant yield of particles relative to the reaction plane:
\begin{equation}
    E\frac{d^{3}N }{d^{3}p}=\frac{1}{2\pi }\frac{d^{2}N }{p_{t}dp_{t}dy}(1+\sum_{n=1}^{\infty } 2v_{n}cos[n(\phi -\Psi_{r})])
\label{equ:v2-psi}
\end{equation}
The second coefficient $v_{2}=<cos[2(\phi-\Psi_{r})]>$ corresponds to the elliptic flow. 
Note that for AMPT events, the flow effect was realized with the orientation of reaction plane $\Psi_{r}$ fixed to be zero by default.

As was mentioned in previous sections, the centrality intervals at AMPT to measure the event-averaged $v_{2}$ were organized in the same percentage groups as the STAR CME work~\cite{PhysRevC.105.014901} for better comparison.
We then continued to look for possible bias of $v_{2}$ in multiple combinations of particle selection in their charge properties, $\eta $ \,and \pt, when determining centrality and mean multiplicity.

The $v_{2}$ methods were also varied for this purpose, not only limited to the preliminary reaction plane way, but also those more complex methods applied at STAR work, such as 2-particle correlation and multi-sub-detector event gap.

$\bullet$ The ideal reaction plane method $v_{2real}=<<cos(2\phi) - \Psi_r>>$ was listed here in the beginning. In this preliminary definition, $\phi$ was azimuth angle for each particle we take into account the $v_{2}$ background, and their \pt\ ranges from 0.2 to 2 \gevc, $| \eta |<1$ to simulate the TPC acceptability as STAR' choice. $<<...>>$ indicates first an average overall particles in the same event, then averaging over all events in that matching centrality interval. To simplify the case, the weight for all particles and all event was set to be one.
Since the $\Psi_r$ in Equ.~\ref{equ:v2-psi} was fixed to be zero at original AMPT, this definition was further simplified to $v_{2real}=<<cos(2\phi)>>$. 

$\bullet$  A more realistic reaction plane method, $v_{2obs}=<<cos(2(\phi-\Psi_{2,PP}))>> / (resolution\ of\ \Psi)$. 
Instead of the fixed $\Psi_r = 0 $ as original AMPT, this $\Psi_{2,PP}$ was rebuilt by the 2nd-order participant plane angle in every event,
\begin{equation}
\Psi_{2,PP} =(tan^{-1}\frac{\sum_{i}^{}sin(2\phi_{i} ) }{\sum_{i}^{}cos(2\phi_{i})})/2 
\label{equ:v2-PP}
\end{equation}
Here $\phi_{i}$ was azimuth angle for particles taken into account as $\Psi_{PP}$ calculation in each event, and was generally different from the $\phi$ of particles chosen for $v_{2}$ background in Equ.~\ref{equ:v2-psi} or \ref{equ:v2-PP} to avoid auto-correlation.

In this analysis, we varied different $\eta$ ranges for particles of $\phi_{i}$ to rebuild $\Psi_{2,PP}$,
then varing the \pt\  ranges from 0.2 to 2\, \gevc,$| \eta |<1$  for particles of $\phi$, but \pt\ $>0.2$\,\gevc\, similar as STAR selections.
All weights of particles and events were also set to be 1.
Meanwhile, the second harmonic event plane resolution of $\Psi_{2,PP}$ was simplified to be $<cos[2(\Psi_{2,PP} - \Psi_r)]>$, the averaged difference between rebuilt $\Psi_{2,PP}$ and the "real" $\Psi_r$. 
It would be more complex for real experiments, when $\Psi_r$ was not directly measurable and resolution of $\Psi$ was usually calculated between $\Psi_{2,PP}$ from multiple sub-detectors of different granularity.

\begin{figure}[h!]
\centering
\includegraphics[width=0.8\linewidth]{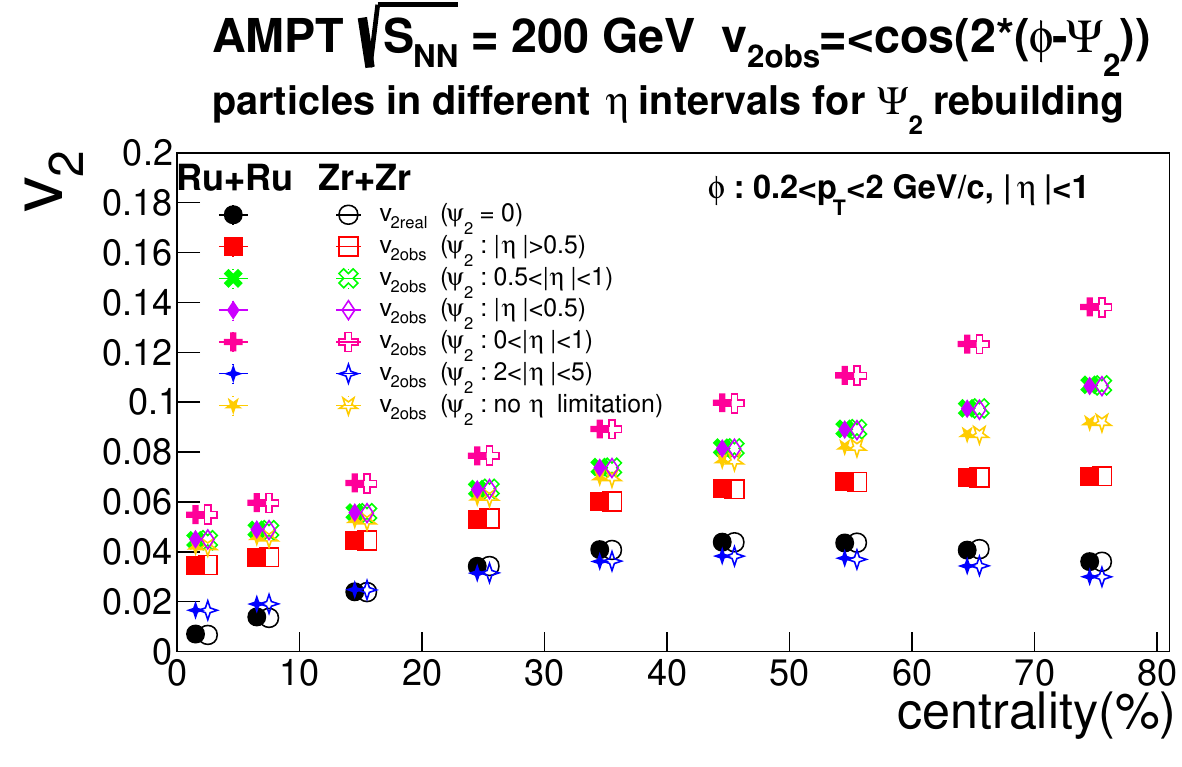}
\caption{
Elliptic anisotropy flow using different reaction-plane methods in isobar collisions at 200 \GeV\ as a function of centrality by AMPT. 
Inside \{\} were the pseudorapidity acceptance of the final state particles used in the reconstruction of $\Psi_{2,PP}$.
}
\label{fig:v2_integral_etacut0.5_different_pp}
\end{figure}

Fig.~\ref{fig:v2_integral_etacut0.5_different_pp} showed all the $v_{2}$ from preliminary reaction-plane methods above.
In this figure, the selection of particle (pseudo-)rapidity intervals for $\Psi_{2,PP}$ clearly resulted in different distribution trends for $v_{2obs}$. 
For more direct comparison with the STAR CME work later, we defined $v_2$\{EP\} by selecting from the $v_{2obs}$ above that of $\phi_{i}$ particles at same $|\eta| > 0.5$ as STAR did, maintaining the balance between larger acceptance and less auto-correlation.


In the next step, two flow methods applied at STAR CME work called 2-particle azimuthal correlations ($v_2$\{2\}) and multi-sub-detector event-gap flow ($v_{2}$\{SP\}(TPC-EPD)), were reproduced at our AMPT analysis to search for possible centrality bias in more details.

$\bullet$ The $v_2$\{2\} of 2-particle azimuthal correlations methods was introduced to make synthetic use of particles for flow and for reaction plane.
\begin{equation}
<cos(n(\phi_{1}-\phi_{2}))>=<e^{in(\phi_{1}-\phi_{2})}>=<v_{n}^{2}>+\delta _{n}
\label{equ:v2-2-particle}
\end{equation}
where $\phi_1$ and $\phi_2$ were all particles selected, and their \pt\ ranges were selected from 0.2 to 2\, \gevc. The $\delta_{n}$ accounted for the non-flow contribution such as jets, and can be subtracted from flow measurements order by order based on multi-particle cumulants.
These cumulants were defined in terms of flow moment vectors $Q_{n}$
\begin{equation}
Q_{n}\equiv \sum_{i=1}^{M}e^{in\phi_{i} }
\label{equ:Qn}
\end{equation}
$M$ was the number of particles selected in each event for flow calculation and $\phi_i$ were their azimuth angles.
So we have $Q_{n}$ written further as
\begin{equation}
|Q_{n}|^2=\sum_{i,j=1}^{M} {e^{in(\phi_{i}-\phi_{j})}}=M+\sum_{i,j=1,i\ne j}^{M}  e^{in(\phi_{i}-\phi_{j})}
\label{equ:Qn-square}
\end{equation}

The single-event averaged 2-particles azimuthal correlations were then defined as 
\begin{equation}
<2>=\frac{|Q_{n}|^2-M}{M(M-1)} 
\label{equ:2-particle-correlation}
\end{equation}
then the averaged 2-particle correlation over all events in the matching centrality 
\begin{equation}
<<2>>\equiv <<e^{in(\phi_{1}-\phi_{2})}>>\equiv \frac{\sum_{events}^{}(W_{<2>})_{i}<2>_{i} }{\sum_{events}^{}(W_{<2>})_{i}}
\label{equ:2-particle-correlation-all-events}
\end{equation}
$W_{<2>}=M(M-1)$ was the event weight.
This brought to us, finally, the $v_2\{2\}=\sqrt{<<2>>}$.

To reach best uniform acceptance, the pseudorapidity acceptance for particles contributing to $\phi_i$ and calculating the corresponding particle number M was set to be $|\eta| < 1$ as the STAR TPC detector acceptance in CME work.
Then to guarantee less auto-correlation, we required $|\Delta \eta| > 1$ between each pair of particles for the cumulants of $Q_n$, 
i.e. one particle within $-1<\eta<-0.5$ to calculate $Q_{nL}$ and the other within $0.5<\eta<1$ to calculate $Q_{nR}$, and each corresponds to $M_{L}$ and $M_{R}$ respectively.
Then the whole-event correlation
\begin{equation}
<2>=\frac{Q_{nL}\times Q_{nR}^{*} }{M_{L}\times M_{R} }
\label{equ:2-particle-correlation-LR}
\end{equation}
where was the number of particles within $-1<\eta<-0.5$ as $M_{L}$ and $0.5<\eta<1$ as $M_{R}$. 
Note the star markers at $Q_n$ meaning the conjugate numbers.
The remaining intra-/inter- event averaging operations for $<<2>>$ remained the same as previous $v_2$\{EP\} method. 

$\bullet$ The $v_{2}$\{SP\}(TPC-EPD) of multi-sub-detector event-gap method was also defined for more direct comparison with the STAR results.
Three $Q_{n, detector}$ from sub-detector TPC ($|\eta| < 1$), EPD-East ($2.1<\eta<5.1$), EPD-West ($-5.1<\eta<-2.1$) were calculated first as previous method, each selecting particles within their own $\eta$, and \pt\ between 0.2 to 2\, \gevc \, as STAR.
The weights of $Q_n$ inside each event were defined by the number of particles selected in that sub-detector.

%
\begin{equation}
v_{2}\{\mathrm{SP}\}\mathrm{(TPC-EPD)} =\frac{Q_{n,\mathrm{TPC}}Q_{n,\mathrm{EPD-East}}^{*}+Q_{n,\mathrm{TPC}}Q_{n,\mathrm{EPD-West}}^{*} }{2\sqrt{<Q_{n,\mathrm{EPD-East}}Q_{n,\mathrm{EPD-West}}^{*}>}}
\label{equ:v2-TPC-EPD}
\end{equation}
%

\begin{figure}[h!]
\centering
\includegraphics[width=0.8\linewidth]{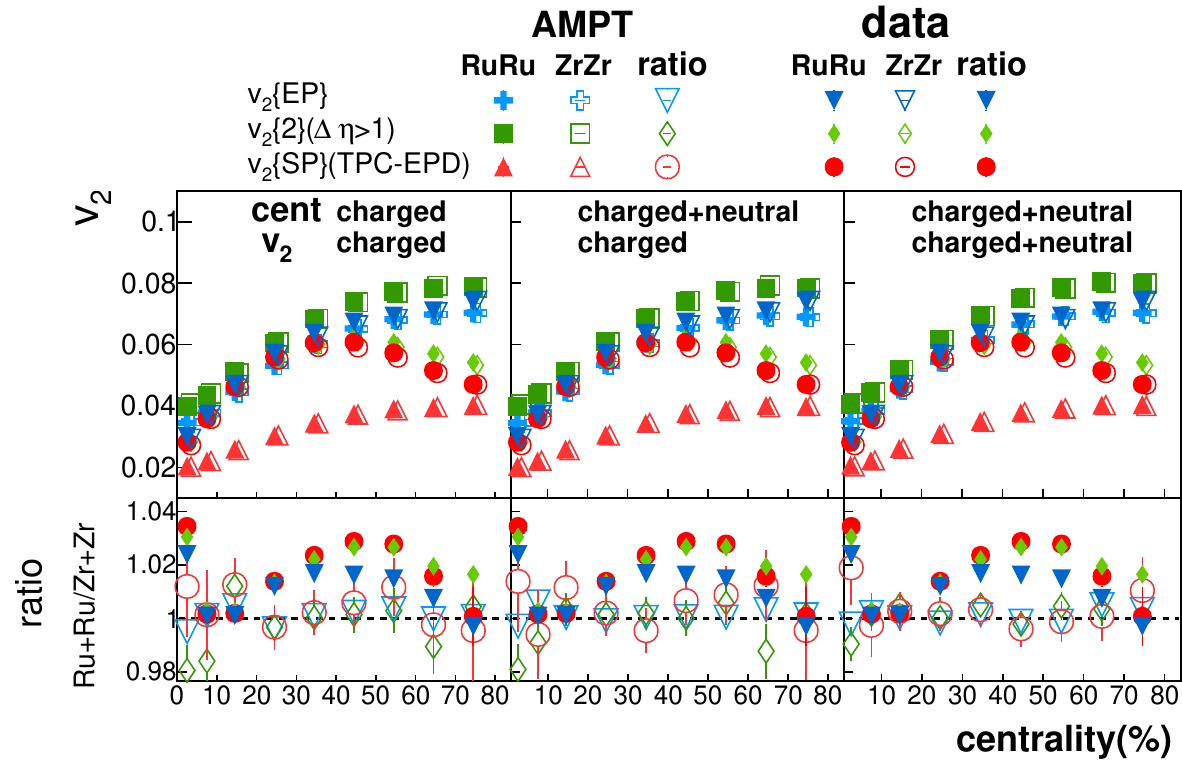}
\caption{
In each panel, the flow of Ru+Ru to Zr+Zr and their ratios from different $v_2$ methods were compared with charged-only STAR results.
The left, middle, and right column panels in the figures correspond to the three combinations of centrality and flow in their electrical properties of particle selection, using the integral centrality method. 
}
\label{fig:v2_integral_etacut0.5_different_particle}
\end{figure}

\begin{figure}[h!]
\centering
\includegraphics[width=0.8\linewidth]{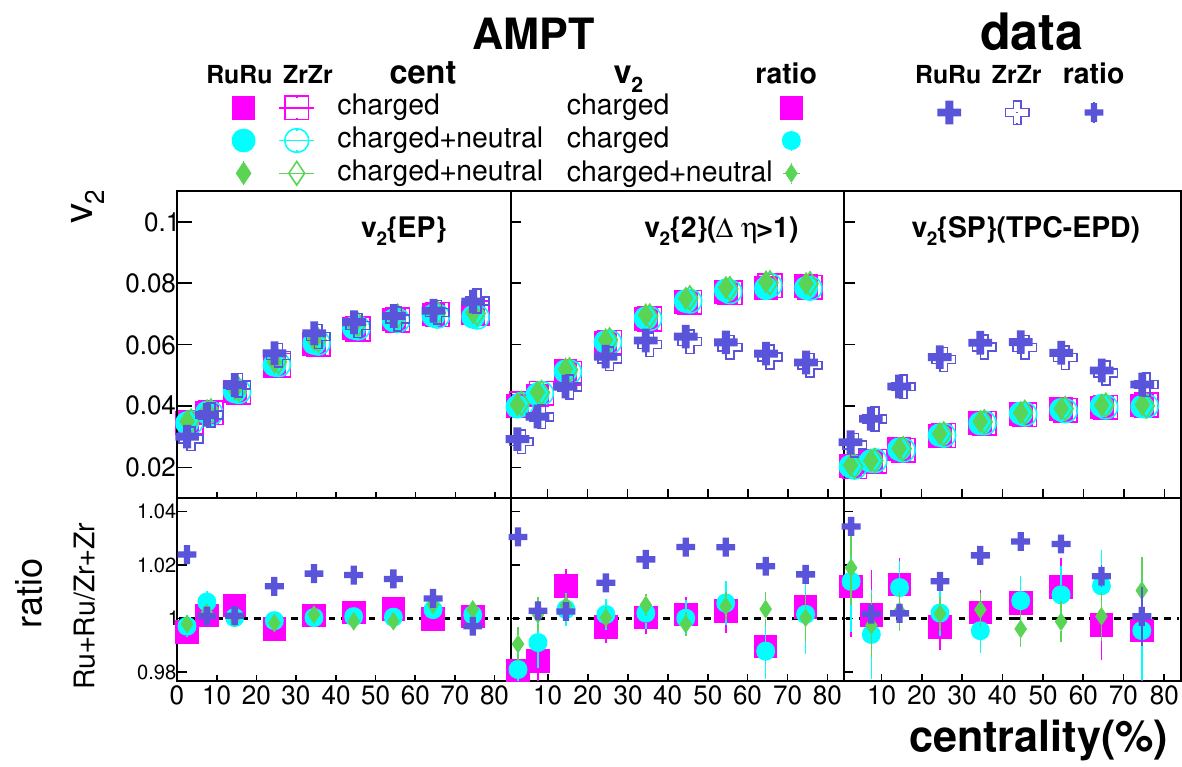}

\caption{
In each panel, the flow of Ru+Ru to Zr+Zr and their ratios from different hadrons of electrical properties in calculating centrality and flow were compared with charged-only STAR results, by reorganizing the points at Fig.~\ref{fig:v2_integral_etacut0.5_different_particle}.
The left, middle, and right column panels in the figures correspond to the $v_2\{EP\}$, $v_2\{2\}$ and $v_{2}\{\mathrm{SP}\} (\mathrm{TPC-EPD})$ method, respectively.
}
\label{fig:v2_integral_etacut0.5_different_v2method}
\end{figure}

\begin{figure}[h!]
\centering
\includegraphics[width=0.8\linewidth]{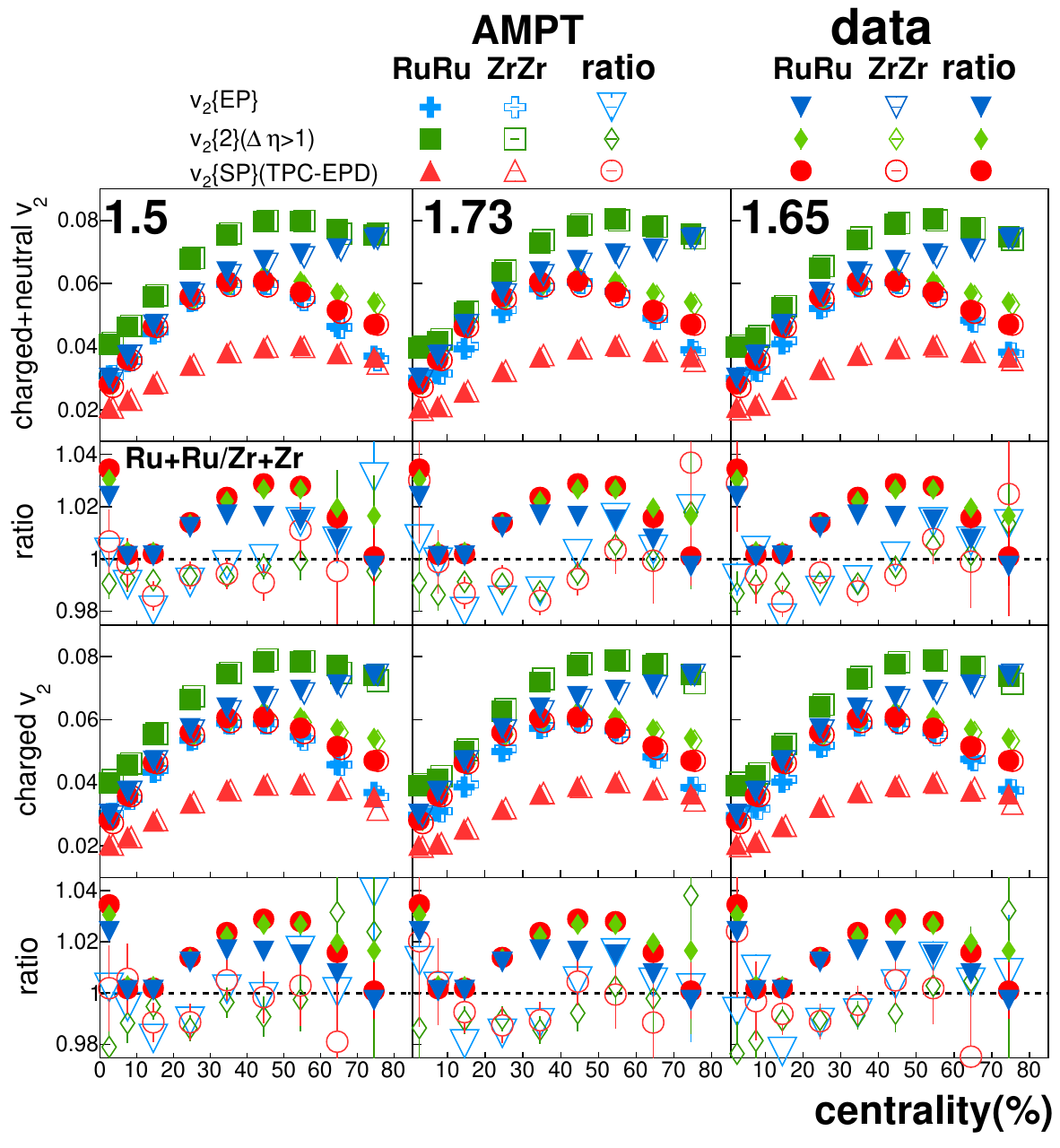}
\caption{
In each panel, the flow of Ru+Ru to Zr+Zr and their ratios from different $v_2$ methods were compared with charged-only STAR results.
The left, middle, and right column panels in the figures correspond to the three coefficients of charged+neutral to charged particles in Tab.~\ref{tab:2}, when STAR multiplicities were applied to the designation centrality method.
The top and bottom panels correspond to whether charged+neutral or charged-only particles were selected in the calculation of $v_2$, respectively.
}
\label{fig:v2_designated_etacut0.5_different_multiply}
\end{figure}

\begin{figure}[h!]
\centering
\includegraphics[width=0.8\linewidth]{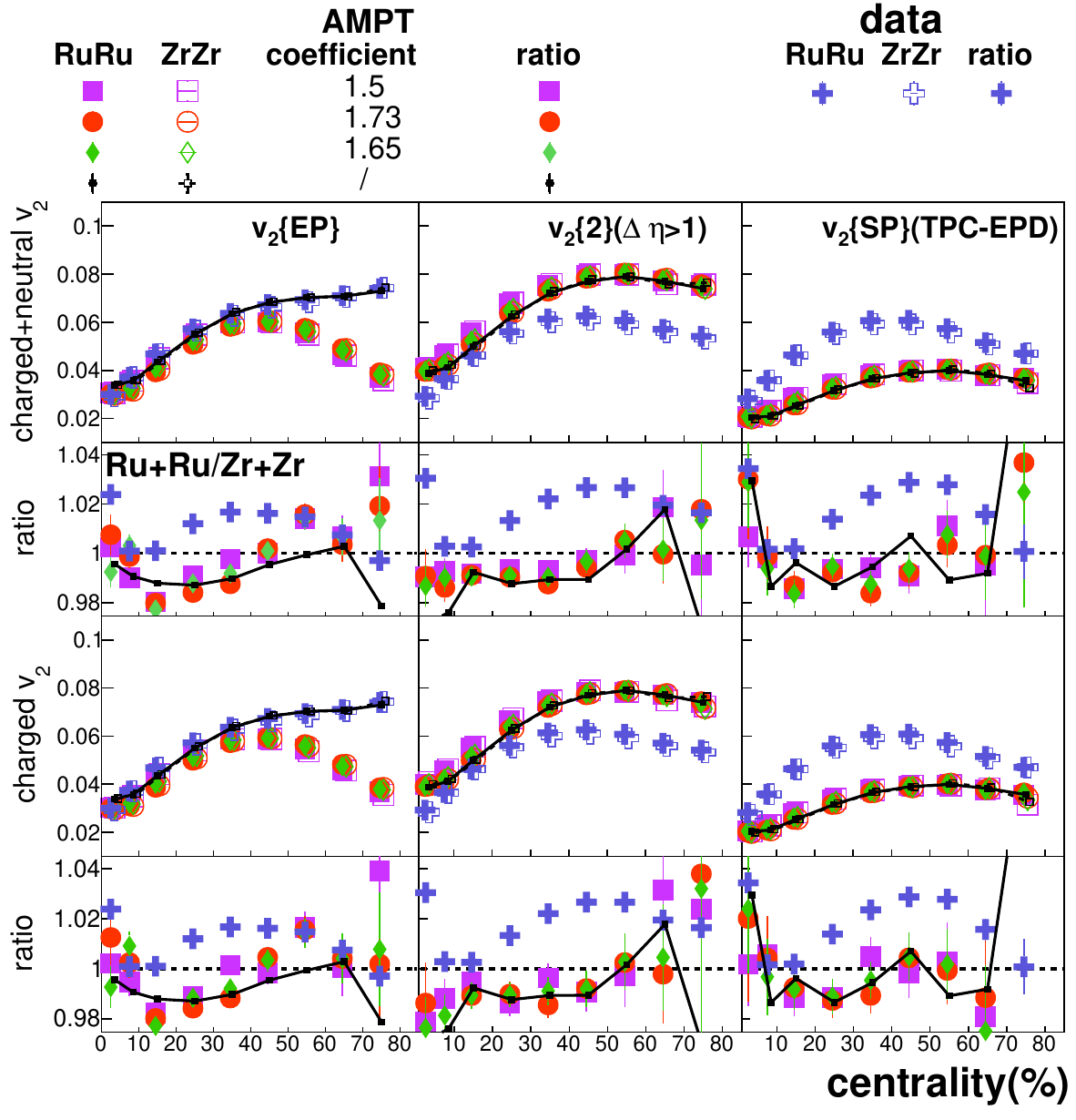}

\caption{
In each panel, based on the designation centrality method, the flow of Ru+Ru to Zr+Zr and their ratios from different centrality coefficients were compared with charged-only STAR results.
The AMPT results with centrality based on charged particles only were also plotted as solid lines.
The left, middle, and right columns corresponded to the three $v_2$ methods.
The top and bottom panels correspond to whether charged+neutral
or charged-only particles were selected in the calculation of $v_2$, respectively.
}
\label{fig:v2_designated_etacut0.5_different_multiply_v2method}
\end{figure}

The detailed comparison of these $v_2$ methods between AMPT and STAR were shown in Fig.~\ref{fig:v2_integral_etacut0.5_different_particle}, \ref{fig:v2_integral_etacut0.5_different_v2method}, \ref{fig:v2_designated_etacut0.5_different_multiply},
and \ref{fig:v2_designated_etacut0.5_different_multiply_v2method},
with multiple combinations of particles selections for either centrality or $v_2$ at AMPT.
They will be discussed in the next section.


\section{Discussion}
\label{sec:discussion}

As mentioned above, the main purpose of this analysis was to explore possible bias in the measurements of the flow background at isobar experiments of $_{44}^{96}Ru+_{44}^{96}Ru$ and $_{40}^{96}Zr+_{40}^{96}Zr$ at \snn\ = 200 \GeV~\cite{PhysRevC.105.014901},
one of whose basic assumptions was that the same number of initial colliding nucleons produces the same number of final-state hadrons for centrality and flow background.
Meanwhile, the STAR experiment often selected only charged hadrons for centrality and flow measurements for better acceptance and momentum resolution.
Therefore, we studied in AMPT the effect of charge property,
by picking up identical nuclear structure parameters for \ruru\ and \zrzr\,
and found that the yields of charged final-state hadrons at the same centrality without $\eta$ limit were slightly higher in \ruru\ than in \zrzr\ as Tab.\ref{tab:5} vs. \ref{tab:6} showed.

This was probably related to the fact that Ru has more protons and the charge conservation at AMPT, 
and might indicate a possible bias in STAR centrality and/or flow measurements if only charged particles were considered.
However, such bias could reduce to a much smaller size within the STAR TPC detector acceptance of $|\eta| < 0.5$, 
as AMPT showed in this $\eta$ limit there was basically no difference in the number of final-state hadrons between \ruru\ and \zrzr\ in Tab.\ref{tab:3} vs. \ref{tab:4}.

We were then inspired to study in the AMPT model the inclusive final-state hadrons, both charged and neutral, of their contributions to the centrality and flow.
We investigated centrality definition of different hadrons selection of charge property,
about multiplicity distributions, mean multiplicities and their ratios at pseudorapidity $|\eta| < 0.5$,
then calculated $v_{2}$ and their ratios also varying hadron selections.
Since the charge-to-nucleon ratio was different between \ruru\ and \zrzr\ nuclei, we had expected that if the STAR flow background showed any isobar or centrality dependence when only charged hadrons were considered,
it would change quantitatively when the charged+neutral hadrons were included.
 
The results were first presented at Fig.~\ref{fig:v2_integral_etacut0.5_different_particle}, where for each of three $v_2$ methods, 
the AMPT results applied the integration method of centrality as Sec.~\ref{sec:methods},
then calculated $v_2$ using methods at Sec.~\ref{sec:results}.
Both centrality and $v_2$ chose different hadron selections (electrical property) from left to right column of Fig.~\ref{fig:v2_integral_etacut0.5_different_particle},
also compared with the STAR work.
The ratio of flow between \ruru\ and \zrzr\ were further calculated and compared with STAR in the lower panels.
Note the STAR data only used charged particles, thus being identical from left to right columns.

In this figure, unlike the STAR data where centrality dependence of the flow ratios showed,
especially a peak above unity at mid-central collisions where the magnetic effect had been expected to maximize, 
the AMPT results presented a much slighter impact of centrality, regardless of $v_2$ method or electrical property of particles.
The ratio between \ruru\ and \zrzr\ stuck to unity within one percent universally for AMPT, in particularly at the mid-central collisions, in all three lower panels.
This seems to indicate that the hadron selections of charge property, in either centrality or $v_2$ definition, make trivial effect in the decision of relative flow background.

The comparison of charge property between AMPT vs. STAR were reorganized at Fig.~\ref{fig:v2_integral_etacut0.5_different_v2method},
where each column of panels presented flow and their ratios via a different $v_2$ method.
In the left column, the selected $v_2$\{EP\} of AMPT were consistent with STAR within uncertainties as validation of the preliminary event-plane method,
but their ratios showing much less centrality dependence than STAR, as Fig.~\ref{fig:v2_integral_etacut0.5_different_particle} already presented.
Then in the middle and right columns, the flow values of AMPT jumped above or below those of STAR for $v_2$\{2\} and $v_{2}$\{SP\}(TPC-EPD), respectively.
This fact was not surprising since we didn't specifically tune AMPT to agree with STAR in particle spectra or flow values.

Instead, we were more interested in the possible bias in hadron selection to flow ratios, which could affect CME measurements in real data.
However, such bias was at max one percent level at the mid-central collisions for all three $v_2$ methods, among all selections of electrical properties.
On the other hand, when approaching the $0\sim5\%$ centrality for the most central collisions, 
the flow ratios obtained from the three $v_2$ methods at AMPT show three opposite trends at each column from left to right: staying at, dropping below, or rising above unity.
But the uncertainties of AMPT points were not adequate to reach further conclusion at the current stage.

As a complement of flow analysis, the designation centrality method at Sec.~\ref{sec:methods} was also studied.
To explore possible bias in hadron selection, the STAR multiplicities based on charged-only particles were transformed to the centrality interval thresholds in AMPT using all kinds of inclusive-to-charged coefficients from Tab.~\ref{tab:2}.
Figure~\ref{fig:v2_designated_etacut0.5_different_multiply} presented AMPT vs. STAR in $v_2$ and their ratios as functions of centrality, based on the coefficients of 1.5 from naive fragmentation, 1.73 from hadrons, and 1.65 from pions in the left, middle and right columns, respectively.
The particles used to calculate $v_2$ also varied from charged+neutral at top panels to charged only at bottom panels.
In each panel, the corresponding AMPT results from that centrality and $v_2$ particle selection was compared with STAR data of all three $v_2$ methods.
STAR data used only charged particles and kept identical from left to right columns.

Similar as the integral centrality method at Fig.~\ref{fig:v2_integral_etacut0.5_different_particle} and \ref{fig:v2_integral_etacut0.5_different_v2method},
the designation centrality method at Fig.~\ref{fig:v2_designated_etacut0.5_different_multiply} 
also showed trivial centrality dependence of $v_2$ ratios within one percent, if any, in all combinations of centrality coefficients and $v_2$ methods, in contrary to the STAR work at mid-central collisions.
Furthermore, no evident difference was found at the flow ratios between charged+neutral and charged-only hadrons or various $v_2$ methods.

As did previously, the results from designation centrality method were also reorganized to sort by $v_2$ methods in Fig.~\ref{fig:v2_designated_etacut0.5_different_multiply_v2method}.
In each column, the AMPT flow and their ratios of multiple centrality coefficients were compared with STAR data of the same $v_2$ method.
In this figure, we also presented AMPT results using only charged particles for both centrality and $v_2$, same as STAR did in their CME work, as solid lines.
For better visual effect, the flow ratios were zoomed in and the most peripheral points might be out-of-scope.
While the AMPT points not necessarily being consistent with STAR, the relative difference among flow ratios when different coefficients were chosen mostly stayed within one percent, especially in the mid-central centrality.

In conclusion, the relative difference among AMPT results, in either integration or designation centrality methods, suggested the bias of selecting only charged particles in STAR was constrained to be with one percent, assuming that AMPT reflected the particle distribution in a high consistency with it.
Meanwhile, if the real experiments can further upgrade the detectors to make full use of the neutral particles, it may provide better understanding of the relative difference between the flow background of the isobar experiments, and constrain their contribution to the CME signal in a better-defined regime.



\section{ACKNOWLEDGMENTS}

We thank Y.X. Mao for the helpful discussions. This work was supported by Supported by National Natural Science Foundation of China under Grant No. 12061141008, and National Key R\&D Plan of China under Grant No. 2024YFA1611003.

\bibliographystyle{unsrt}
\bibliography{main}

\begin{thebibliography}{10}

\bibitem{PhysRevC.105.014901}
{Abdallah, M. S. and Aboona, B. E. and Adam, J. et al.}
\newblock {Search for the chiral magnetic effect with isobar collisions at $\sqrt{{s}_{NN}}=200$ GeV by the STAR Collaboration at the BNL Relativistic Heavy Ion Collider}.
\newblock {\em Phys. Rev. C}, 105:014901, Jan 2022.

\bibitem{PhysRevC.72.064901}
Zi-Wei Lin, Che~Ming Ko, Bao-An Li, Bin Zhang, and Subrata Pal.
\newblock Multiphase transport model for relativistic heavy ion collisions.
\newblock {\em Phys. Rev. C}, 72:064901, Dec 2005.

\bibitem{PhysRevC.85.044907}
Wei-Tian Deng and Xu-Guang Huang.
\newblock Event-by-event generation of electromagnetic fields in heavy-ion collisions.
\newblock {\em Phys. Rev. C}, 85:044907, Apr 2012.

\bibitem{BZDAK2012171}
Adam Bzdak and Vladimir Skokov.
\newblock Event-by-event fluctuations of magnetic and electric fields in heavy ion collisions.
\newblock {\em Physics Letters B}, 710(1):171--174, 2012.

\bibitem{BLOCZYNSKI20131529}
John Bloczynski, Xu-Guang Huang, Xilin Zhang, and Jinfeng Liao.
\newblock Azimuthally fluctuating magnetic field and its impacts on observables in heavy-ion collisions.
\newblock {\em Physics Letters B}, 718(4):1529--1535, 2013.

\bibitem{PhysRevC.70.057901}
Sergei~A. Voloshin.
\newblock Parity violation in hot qcd: How to detect it.
\newblock {\em Phys. Rev. C}, 70:057901, Nov 2004.

\bibitem{2006.05035}
Abdallah et~al.
\newblock Pair invariant mass to isolate background in the search for the chiral magnetic effect in au+au collisions at 200gev.
\newblock {\em Physical Review C}, 106(3), sep 2022.

\bibitem{PhysRevC.97.044912}
Sirunyan et~al.
\newblock Constraints on the chiral magnetic effect using charge-dependent azimuthal correlations in $p\mathrm{Pb}$ and pbpb collisions at the cern large hadron collider.
\newblock {\em Phys. Rev. C}, 97:044912, Apr 2018.

\bibitem{PhysRevC.81.064902}
Fuqiang Wang.
\newblock Effects of cluster particle correlations on local parity violation observables.
\newblock {\em Phys. Rev. C}, 81:064902, Jun 2010.

\bibitem{PhysRevC.81.031901}
Adam Bzdak, Volker Koch, and Jinfeng Liao.
\newblock Remarks on possible local parity violation in heavy ion collisions.
\newblock {\em Phys. Rev. C}, 81:031901, Mar 2010.

\bibitem{PhysRevC.83.014913}
S\"oren Schlichting and Scott Pratt.
\newblock Charge conservation at energies available at the bnl relativistic heavy ion collider and contributions to local parity violation observables.
\newblock {\em Phys. Rev. C}, 83:014913, Jan 2011.

\bibitem{PhysRevC.101.034912}
Jie Zhao, Yicheng Feng, Hanlin Li, and Fuqiang Wang.
\newblock hijing can describe the anisotropy-scaled charge-dependent correlations at the bnl relativistic heavy ion collider.
\newblock {\em Phys. Rev. C}, 101:034912, Mar 2020.

\bibitem{Nucl.Sci.Tech.32.113}
Zi-Wei Lin and Liang Zheng.
\newblock Further developments of a multi-phase transport model for relativistic nuclear collisions.
\newblock {\em Nuclear Science and Techniques}, 32, Oceb 2021.

\bibitem{PhysRevD.44.3501}
Xin-Nian Wang and Miklos Gyulassy.
\newblock hijing: A monte carlo model for multiple jet production in $\mathrm{pp}$, $\mathrm{pA}$, and $\mathrm{AA}$ collisions.
\newblock {\em Phys. Rev. D}, 44:3501--3516, Dec 1991.

\bibitem{GYULASSY1994307}
Miklos Gyulassy and Xin-Nian Wang.
\newblock Hijing 1.0: A monte carlo program for parton and particle production in high energy hadronic and nuclear collisions.
\newblock {\em Computer Physics Communications}, 83(2):307--331, 1994.

\bibitem{ZHANG1998193}
Bin Zhang.
\newblock Zpc 1.0.1: a parton cascade for ultrarelativistic heavy ion collisions.
\newblock {\em Computer Physics Communications}, 109(2):193--206, 1998.

\bibitem{JGXK2024090101}
Si-Yu Tang, Liang Zheng, Xiao-Ming Zhang, and Ren-Zhuo Wan.
\newblock Investigating the elliptic anisotropy of identified particles in p–pb collisions with a multi-phase transport model.
\newblock {\em Nuclear Science and Techniques}, 35(2):32, 2024.

\bibitem{PhysRevC.52.2037}
Bao-An Li and Che~Ming Ko.
\newblock Formation of superdense hadronic matter in high energy heavy-ion collisions.
\newblock {\em Phys. Rev. C}, 52:2037--2063, Oct 1995.

\bibitem{PhysRevC.94.041901}
Wei-Tian Deng, Xu-Guang Huang, Guo-Liang Ma, and Gang Wang.
\newblock Testing the chiral magnetic effect with isobaric collisions.
\newblock {\em Phys. Rev. C}, 94:041901, Oct 2016.

\bibitem{PhysRevLett.127.242301}
Giuliano Giacalone, Jiangyong Jia, and Chunjian Zhang.
\newblock Impact of nuclear deformation on relativistic heavy-ion collisions: Assessing consistency in nuclear physics across energy scales.
\newblock {\em Phys. Rev. Lett.}, 127:242301, Dec 2021.

\bibitem{SHOU2015215}
Q.Y. Shou, Y.G. Ma, P.~Sorensen, A.H. Tang, F.~Videbæk, and H.~Wang.
\newblock Parameterization of deformed nuclei for glauber modeling in relativistic heavy ion collisions.
\newblock {\em Physics Letters B}, 749:215--220, 2015.

\bibitem{XU2021136453}
Hao jie Xu, Hanlin Li, Xiaobao Wang, Caiwan Shen, and Fuqiang Wang.
\newblock Determine the neutron skin type by relativistic isobaric collisions.
\newblock {\em Physics Letters B}, 819:136453, 2021.

\bibitem{ACKERMANN1999681}
K.H.~Ackermann et~al.
\newblock The star time projection chamber.
\newblock {\em Nuclear Physics A}, 661(1):681--685, 1999.

\bibitem{PhysRevLett.128.022301}
Chunjian Zhang and Jiangyong Jia.
\newblock Evidence of quadrupole and octupole deformations in $^{96}\mathrm{Zr}+^{96}\mathrm{Zr}$ and $^{96}\mathrm{Ru}+^{96}\mathrm{Ru}$ collisions at ultrarelativistic energies.
\newblock {\em Phys. Rev. Lett.}, 128:022301, Jan 2022.

\end{thebibliography}

\end{document}